\documentclass[fleqn,usenatbib]{mnras}

\usepackage{newtxtext,newtxmath}

\usepackage[T1]{fontenc}
\usepackage{ae,aecompl}
\usepackage{xcolor}
\usepackage{ulem}

\usepackage{graphicx}	
\usepackage{amsmath}	
\usepackage{amssymb}	

\def\ie{{\it i.e.}}

\title[I-fronts around moving BHs]{Structure and Instability of the Ionization Fronts around Moving Black Holes}

\author[K. Sugimura and M. Ricotti]{
  Kazuyuki Sugimura$^{1}$\thanks{E-mail: sugimura@astro.umd.edu}
  and
  Massimo Ricotti$^{1}$\thanks{E-mail: ricotti@astro.umd.edu}
  \\
  $^{1}$Department of Astronomy, University of Maryland, College Park, MD 20740, USA
  }

  \date{Accepted XXX. Received YYY; in original form ZZZ}

  \pubyear{2020}

  \hypersetup{draft} 
  
\begin{document}
  \label{firstpage}
  \pagerange{\pageref{firstpage}--\pageref{lastpage}}
  \maketitle

  \begin{abstract}
In this paper we focus on understanding the physical processes that lead
to stable or unstable ionization fronts (I-fronts) observed in
simulations of moving black holes (BHs). The front instability may
trigger bursts of gas accretion, rendering the BH significantly more
luminous than at steady-state. We perform a series of idealized three
dimensional radiation hydrodynamics simulations resolving the I-fronts
around BHs of mass $M_\mathrm{BH}$ and velocity $v_\infty$ accreting
from a medium of density $n_\mathrm{H}$.  The I-front, with radius
$R_\mathrm{I}$, transitions from D-type to R-type as the BH velocity
becomes larger than a critical value $v_\mathrm{R}\sim
40\,\mathrm{km/s}$. The D-type front is preceded by a bow-shock of
thickness $\Delta R_\mathrm{I}$ that decreases as $v_\infty$ approaches
$v_\mathrm{R}$. We find that both D-type and R-type fronts can be
unstable given the following two conditions: i) for D-type fronts the
shell thickness must be $\Delta R_\mathrm{I}/R_\mathrm{I}<0.05$ (\ie,
$v_\infty \gtrsim 20\,\mathrm{km/s}$.), while no similar restriction
holds for R-type fronts; ii) the temperature jump across the I-front
must be $T_\mathrm{II}/T_\mathrm{I}>3$.  This second condition is
satisfied if $T_\mathrm{I}<5000\,\mathrm{K}$ or if
$n_\mathrm{H}\,M_\mathrm{BH} \gtrsim 10^6\,M_\odot\,\mathrm{cm^{-3}}$.
Due to X-ray pre-heating typically $T_\mathrm{I} \sim 10^4\,\mathrm{K}$,
unless the D-type shell is optically thick to X-rays, which also happens
when $n_\mathrm{H}\,M_\mathrm{BH}$ is greater than a
metallicity-dependent critical value.  We thus conclude that I-fronts
around BHs are unstable only for relatively massive BHs moving trough
very dense molecular clouds. We briefly discuss the observational
consequences of the X-ray luminosity bursts likely associated with this
instability.
\end{abstract}




 \section{Introduction}
\label{sec:introduction}

After the merger of galaxies hosting intermediate-mass black holes
($10^2\,M_\odot\lesssim M_\mathrm{BH}\lesssim 10^5\,M_\odot$; IMBHs),
the IMBHs are likely to move around the post-merger galaxies while
accreting the surrounding gas.  The accretion growth of such IMBHs is
considered to play a critical role in the stellar-mass BH seed scenario
of the formation of supermassive BHs ($M_\mathrm{BH}\gtrsim
10^6\,M_\odot$; SMBHs) that reside at the center of almost all galaxies
\citep[see, e.g., ][for review]{Volonteri:2012ab,Inayoshi:2019ab}.
Furthermore, X-ray emission from the accretion disks of IMBHs can build
up an X-ray background that may affect the star-formation history of the
Universe \citep[e.g.,][]{Ricotti:2016ab}. The efficiency of gas
accretion, however, is yet to be fully understood because of the reasons
described below.

Recently, theoretical works have revealed that various factors can
widely change the efficiency of accretion.  In the simplest case of a
stationary BH emitting isotropic radiation, the formation of the ionized
bubble significantly reduces the accretion rate
\citep{Milosavljevic:2009ab,Milosavljevic:2009aa,Park:2011aa,Park:2012aa}.
The reduction, however, is notably weakened if the radiation is
anisotropic as a result of the disk shadowing effect
\citep{Sugimura:2017ab,Takeo:2018aa}, or if the surrounding gas is so
dense as to confine the ionized bubble to the vicinity of the BH
\citep{Inayoshi:2016ac}. Other effects such as the radiation pressure
and attenuation by dust grains \citep{Yajima:2017aa,Toyouchi:2019aa} and
the gas angular momentum \citep{Sugimura:2018aa} are also critical
factors that determine the accretion rate.

For the more realistic case of a BH moving in the surrounding medium (or
equivalently a stationary BH in a homogeneously moving medium), the gas
accretion under radiative feedback was first studied in \citet[hereafter
PR13]{Park:2013aa}. PR13 performed axisymmetric 2D simulations assuming
that gas flows into the computational region from the polar direction of
the spherical coordinates.  They found that the ionization front
(I-front) is D-type and a dense shell forms in front of it in the
low-velocity case, but that it becomes R-type and the shell disappears
in the high-velocity case.  The gravitational attraction by the dense
shell can accelerate the BH, preventing it from sinking to the center of
the galaxy by dynamical friction \citep{Park:2017aa,Toyouchi:2020aa}.
PR13 also found that the accretion rate increases with BH velocity in
the regime when the I-front is D-type, contrary to the classical
Bondi-Hoyle-Lyttleton (BHL) accretion, for which the accretion rate is a
monotonically decreasing function of the velocity.  More interestingly,
the I-front becomes unstable near the D- and R-type transition, and the
instability triggers accretion bursts. The bursty-mode accretion can
enhance the BH growth rate. Moreover, even if the mean accretion rate is
weakly affected by the accretion bursts, the associated luminosity
bursts are significantly brighter than the mean, because the radiative
efficiency of BH accretion disk is higher with the higher accretion rate
in the so-called radiatively inefficient accretion flow (RIAF) regime
\citep{Narayan:1994aa}.

The axisymmetric 2D simulations in PR13, however, are not suitable for
studying the instability of the I-front for two reasons.  First,
although the instability grows along the polar axis in their
simulations, the gas dynamics near the polar axis is likely affected by
numerical artifacts because the polar axis is the coordinate singularity
of the spherical coordinates.  Second, the growth and saturation of the
instability is essentially a 3D process, as we will show in this paper,
and cannot be followed by the 2D simulations.  Therefore, we need to
perform new 3D simulations to reveal the true nature of the instability.

3D simulations that resolve the inner scale where the BH gravity
dominates the gas dynamics (\ie, the scale of the BHL radius, typically
a hundred times smaller than the I-front) can follow the gas accretion
taking into account the interplay of the instability and the modulated
luminosity. The analysis of such simulations, however, is challenging
because of the complicated interactions of various physical processes.
Moreover, the wide dynamic range makes the computational cost expensive
since we require 3D simulations with high resolution near the BH to
resolve feedback effects, and near the I-front to resolve the
instability.

Therefore, as a first step toward understanding the BH accretion rate
and luminosity modulated by the instability of the I-front, we focus on
resolving the region outside the BHL radius by performing a set of 3D
radiation hydrodynamics simulations of gas dynamics near the I-front,
assuming constant BH luminosity and neglecting the BH gravity (since the
I-front radius is much larger than the BHL radius). By extending the
analytical model developed in PR13, we provide a physically motivated
model for the observed properties of the bow shocks in simulated D-type
fronts. We also discuss the conditions for the instability based on the
simulation results.

This paper is organized as follows. In Sec.~\ref{sec:analytical-model},
we describe the analytical model. In Sec.~\ref{sec:simulations} and
Sec.~\ref{sec:results}, we describe our 3D radiation hydrodynamics
simulations and present the simulation results, respectively.  We
discuss the implication of our work in Sec.~\ref{sec:discussion} and
finally present a summary in Sec.~\ref{sec:summary}.

 \section{Analytical model of flow around moving BH with radiation feedback}
\label{sec:analytical-model}

  \subsection{Park-Ricotti model}
\label{sec:1d-model}

Here, we briefly review the analytical model developed in PR13.  This
quasi-1D model of flow around a moving BH with radiation feedback
reproduces the variation of the type of flow and the resulting accretion
rate as a function of BH velocity observed in their simulations. We
refer the readers to PR13 for the details of this model.

The model considers the structure of flow along the axis of the BH
motion. In the BH rest frame, the gas is moving toward the BH with
velocity $v_\infty$.  Hereafter, the subscripts $\mathrm{I}$ and
$\mathrm{II}$ indicate the neutral and ionized sides of the I-front,
respectively.  For simplicity, we assume that the sound speed (or
temperature) discontinuously changes from $c_\mathrm{I}$ in the neutral
gas upstream of the I-front to $c_\mathrm{II}$ in the ionized gas across
the I-front. In this section, we assume that the BH velocity with
respect to the neutral ambient gas is supersonic (\ie, $v_\infty \ge
c_\infty$), and the ambient gas has constant temperature, \ie,
$c_\mathrm{I}=c_\infty$.  We consider the subsonic case ($v_\infty <
c_\infty$), as well as the effect of X-ray pre-heating of the ambient
gas ($c_\mathrm{I} \not= c_\infty$), in
Appendix~\ref{sec:analytical-model-appendix}.

Let us begin with writing down the jump conditions for density $\rho$
and velocity $v$ across the I-front,
\begin{align}
 \frac{\rho_\mathrm{II}}{\rho_\mathrm{I}}
 =\frac{v_\mathrm{I}}{v_\mathrm{II}}
 = \frac{c_\mathrm{I}^2+v_\mathrm{I}^2
 \pm\sqrt{(c_\mathrm{I}^2+v_\mathrm{I}^2)^2-4c_\mathrm{II}^2v_\mathrm{I}^2}}
 {2c_\mathrm{II}^2}
 \quad \Big(\equiv\Delta^{(\pm)} (v_\mathrm{I},\,c_\mathrm{I},\,c_\mathrm{II})\Big)\,.
 \label{eq:jump_IF}
\end{align}
In order for Eq.~\eqref{eq:jump_IF} to have a (real) solution, the
condition $v_\mathrm{I} < v_\mathrm{D}$ or $v_\mathrm{I} > v_\mathrm{R}$
must be satisfied. Here, the D-type critical velocity is defined as
\begin{align}
 v_\mathrm{D}(c_\mathrm{I},\,c_\mathrm{II}) \equiv c_\mathrm{II} - \sqrt{c_\mathrm{II}^2-c_\mathrm{I}^2}
 \approx \frac{c_\mathrm{I}^2}{2c_\mathrm{II}}
 \quad (c_\mathrm{I} \ll c_\mathrm{II}) \,,
 \label{eq:vD}
\end{align}
and the R-type critical velocity is
\begin{align}
 v_\mathrm{R}(c_\mathrm{I},\,c_\mathrm{II}) \equiv c_\mathrm{II} + \sqrt{c_\mathrm{II}^2-c_\mathrm{I}^2}
 \approx 2c_\mathrm{II}
 \quad (c_\mathrm{I} \ll c_\mathrm{II}) \,.
 \label{eq:vR}
\end{align}
The I-fronts with $v_\mathrm{I} < v_\mathrm{D}$ and $v_\mathrm{I} >
v_\mathrm{R}$ are called D-type and R-type, respectively.

In the case with $v_\infty > v_\mathrm{R}$, the gas can directly reach
the I-front with $v_\mathrm{I} = v_\infty$ and $\rho_\mathrm{I} =
\rho_\infty$. The I-front becomes weak R-type (\ie, the minus sign is
taken in Eq.~\ref{eq:jump_IF}), and the density and velocity inside the
HII region are given, respectively, as
\begin{align}
 \rho_\mathrm{II,R}  = \Delta^{(-)}(v_\infty,\,c_\mathrm{I},\,c_\mathrm{II})\, \rho_\infty\,,
 \label{eq:rhoII_wR}
\end{align}  
and   
\begin{align}
v_\mathrm{II,R}  =  \frac{1}{\Delta^{(-)}(v_\infty,\,c_\mathrm{I},\,c_\mathrm{II})}\, v_\infty \,.
 \label{eq:vII_wR}
\end{align}     
Here, the density and velocity jump is generally weak because
$\Delta^{(-)}$ takes its maximum value of $2$ $(c_\mathrm{I} \ll
c_\mathrm{II})$ at $v_\infty=v_\mathrm{R}$ and approaches unity as
$v_\infty$ increases.  Hereafter, we name this type of flows as the
R-type flows.

In the case with $(v_\mathrm{D}< c_\mathrm{I}<)\ v_\infty<v_\mathrm{R}$,
however, the jump conditions in Eq.~\eqref{eq:jump_IF} cannot be
satisfied without extra structures.  In this case, a shock is generated
in front of the I-front. The gas experiencing the shock forms a dense
shell between the shock and the I-front. The I-front becomes D-type
because the gas slows down to below $v_\mathrm{D}$ as a result of the
following two effects. Firstly, the isothermal shock decreases the
velocity by $(v_\infty/c_\mathrm{I})^2$, while increasing the density by
the same factor. Secondly, the tangentially diverging motion decelerates
the gas in the shell (we will discuss this process in detail in
Sec.~\ref{sec:shell-thickness}).  In PR13, they further assumed that the
I-front is critical D-type (\ie, $v_\mathrm{I}=v_\mathrm{D}$), and thus
the velocity inside the HII region is given by
\begin{align}
v_\mathrm{II,D}=c_\mathrm{II}\,.
 \label{eq:vII_cD}
\end{align}     
Here, the total pressure $P_\mathrm{tot}=\rho (v^2+c_\mathrm{s}^2)$,
defined as the sum of the ram pressure $P_\mathrm{ram}=\rho v^2$ and the
thermal pressure $P_\mathrm{th}=\rho c_\mathrm{s}^2$, is conserved
across the shock and the I-front. It is also conserved inside the
shocked shell, where the flow is subsonic, because
$P_\mathrm{tot}\approx P_\mathrm{th}$ and $P_\mathrm{th}$ is conserved
due to the approximate pressure equilibrium.  From Eq.~\eqref{eq:vII_cD}
and $P_\mathrm{tot}=\rho_\infty (v_\infty^2+c_\mathrm{I}^2)$, the
density inside the HII region is given by
\begin{align}
\rho_\mathrm{II,D}=\frac{\rho_\infty (v_\infty^2+c_\mathrm{I}^2)}{2c_\mathrm{II}^2}\,.
 \label{eq:rhoII_cD}
\end{align}     
Hereafter, we name this type of flows as the D-type flows.

\begin{figure}
 \centering \includegraphics[width=8cm]{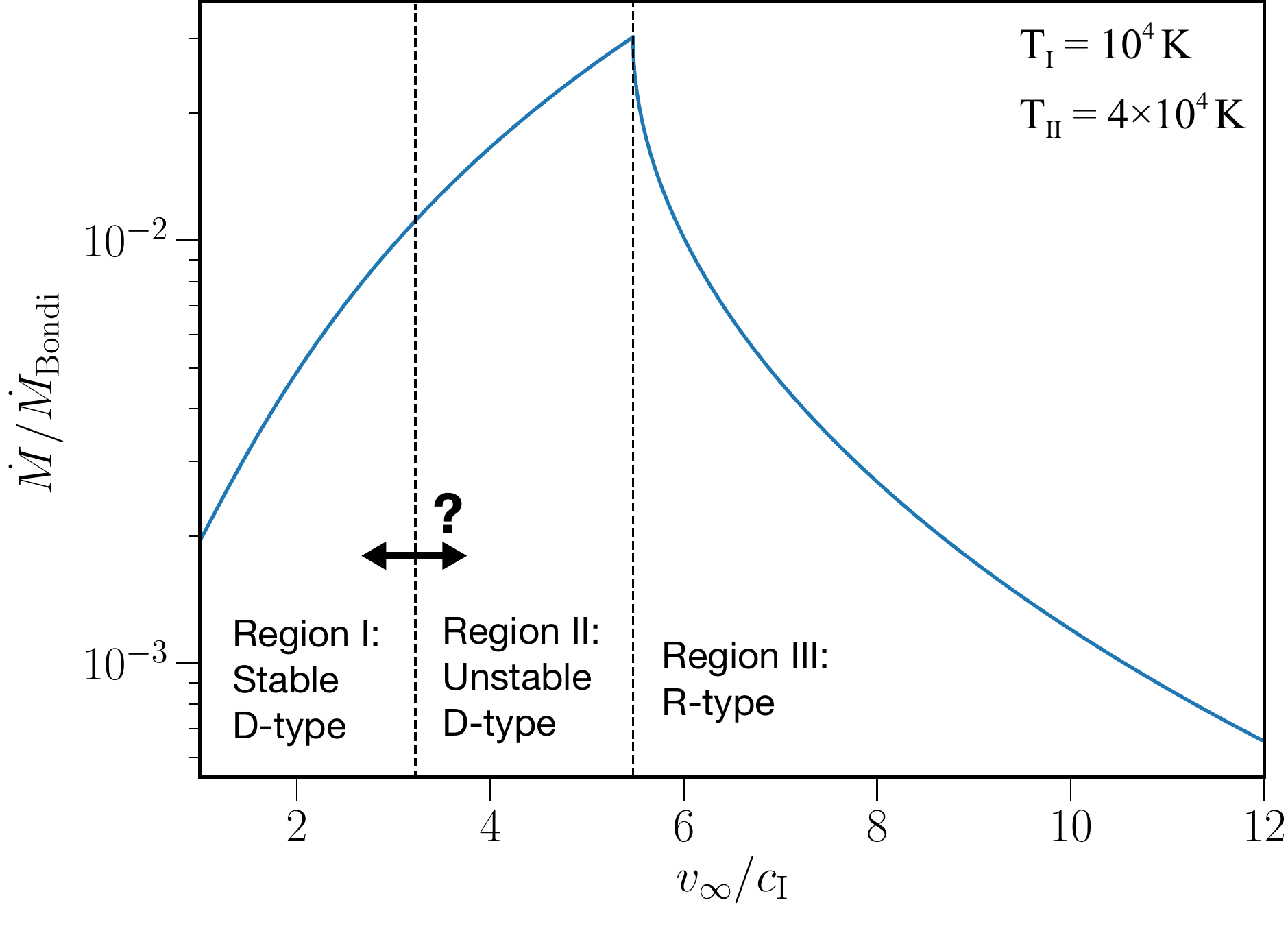}
 \caption{The analytical accretion rate from Park \& Ricotti model.  We
 plot the accretion rate given by Eq.~\eqref{eq:mdot_model} for a BH
 moving at velocity $v_\infty$, with $T_\mathrm{I}=10^4\,\mathrm{K}$ and
 $T_\mathrm{II}=4\times10^4\,\mathrm{K}$.  The accretion rate is
 normalized by the Bondi accretion rate $\dot{M}_\mathrm{Bondi}=4\pi
 \lambda_\mathrm{B} G^2 M_\mathrm{BH}^2/c_\infty^3 $ with
 $\lambda_\mathrm{B}=1$, and thus the dependence on $M_\mathrm{BH}$ and
 $\rho_\infty$ is completely canceled out. There are three regions
 demarcated by the vertical dashed lines: the stable D-type flows
 (left), the unstable D-type flows (center), and the R-type flows
 (right). While the transition from the stable D-type to the unstable
 D-type was seen in the simulations of PR13, the condition for the
 instability was not clear in PR13.}

\label{fig:mdot_PR13}
\end{figure}

Using the density and velocity inside the HII region obtained above, the
accretion rate can be estimated with the Bondi-Hoyle-Lyttleton (BHL)
formula as
\begin{align}
 \dot{M} = \frac{4\pi G^2 M_\mathrm{BH}^2\rho_\mathrm{II} }{(v_\mathrm{II}^2 + c_\mathrm{II}^2)^\frac{3}{2}}
=
\begin{cases}
\dot{M}_\mathrm{D} & c_\mathrm{I}<v_\infty < v_\mathrm{R}\\
\dot{M}_\mathrm{R} & v_\infty > v_\mathrm{R}
\end{cases}
\,,
 \label{eq:mdot_model}
\end{align}
where the accretion rate for the D-type flows
\begin{align}
\dot{M}_\mathrm{D}= \frac{\pi G^2 M_\mathrm{BH}^2\rho_\infty\, (v_\infty^2 + c_\mathrm{I}^2)}{\sqrt{2}c_\mathrm{II}^5}
\label{eq:mdotD}
\end{align}
and that for the R-type flows
\begin{align}
\dot{M}_\mathrm{R}=\frac{4\pi G^2 M_\mathrm{BH}^2\rho_\infty\Delta^{(-)}(v_\mathrm{\infty},c_\mathrm{I},c_\mathrm{II})}
 {\left(c_\mathrm{II}^2+\left(\frac{v_\mathrm{\infty}}{\Delta^{(-)}(v_\mathrm{\infty},c_\mathrm{I},c_\mathrm{II})}\right)^{2}\right)^\frac{3}{2}}
\approx \frac{4\pi G^2 M_\mathrm{BH}^2\rho_\infty}
 {\left(c_\mathrm{II}^2+v_\mathrm{\infty}^2\right)^\frac{3}{2}}  \quad (v_\infty \gg v_\mathrm{R}) \,.
\label{eq:mdotR}
\end{align}
Here, we have used $\rho_\mathrm{II}$ and $v_\mathrm{II}$ given by
Eqs.~\eqref{eq:rhoII_wR} and \eqref{eq:vII_wR} for $\dot{M}_\mathrm{R}$
and those given by Eqs.~\eqref{eq:vII_cD} and \eqref{eq:rhoII_cD} for
$\dot{M}_\mathrm{D}$.  In Fig.~\ref{fig:mdot_PR13}, we plot the
accretion rate given by Eq.~\eqref{eq:mdot_model}, with the temperature
of the neutral gas $T_\mathrm{I}=10^4\,\mathrm{K}$
($c_\mathrm{I}=8.1\,\mathrm{km/s}$) and that of the ionized gas
$T_\mathrm{II}=4\times10^4\,\mathrm{K}$
($c_\mathrm{II}=23\,\mathrm{km/s}$).  We use the same $T_\mathrm{I}$ and
$T_\mathrm{II}$ for the Park \& Ricotti model in the rest of the paper
unless otherwise stated.  As pointed out in PR13, $\dot{M}$ takes its
maximum value at $v_\infty=v_\mathrm{R}$ in Figure~\ref{fig:mdot_PR13}.

For $\dot{M}$ obtained above, the luminosity of the X-ray emission from
the BH accretion disk can be calculated as
\begin{align}
 L=\eta\,\dot{M}\,c^2\,.  
\label{eq:lum_from_acc}
\end{align}
with radiative efficiency $\eta$.  Here, we assume an
$\dot{M}$-dependent efficiency modeled as
\begin{align}
		  \eta = \eta_0\times
\min\left(1,\, \frac{\dot{M}c^2}{L_\mathrm{Edd}}\right)\,,
		  \label{eq:eta_variable}
\end{align}
with the normalization factor $\eta_0=0.1$. The Eddington luminosity is
defined as $L_\mathrm{Edd} \equiv 4\pi G M_\mathrm{BH} c
m_\mathrm{p}/\sigma_\mathrm{T}$, where $m_\mathrm{p}$ is the proton mass
and $\sigma_\mathrm{T}=6.65\times10^{-25}\,\mathrm{cm^{2}}$ is the
Thomson cross-section.  This form of $\eta$ is motivated by the fact
that the radiative efficiency becomes lower with lower $\dot{M}$ in the
in the low accretion rate RIAF regime \citep{Narayan:1994aa}.

In order to better understand the stability of the I-front, we estimate
its size $R_\mathrm{IF}$, using the luminosity and density modeled
above. As pointed out in PR13, $R_\mathrm{IF}$ is either determined by
the balance of the ionization and recombination rates inside the ionized
region, or by the balance of the ionization rate and the flow rate of
neutral gas across the I-front.  In the former case, $R_\mathrm{IF}$ is
well approximated by the Str\"omgren radius,
\begin{align}
r_\mathrm{Strm} \equiv \left(\frac{3\dot{N}_\mathrm{ion}}{4\pi
n_\mathrm{II}^2\alpha_\mathrm{B}}\right)^{\frac{1}{3}} \,,
\label{eq:r_strm}
\end{align}
where $\dot{N}_\mathrm{ion}=L/\left<h\nu\right>$ is the ionizing photon
emission rate, with average energy of ionizing photons
$\left<h\nu\right>$, $n_\mathrm{II}\approx
\rho_\mathrm{II}/m_\mathrm{p}$ is the ionized gas number density, and
$\alpha_\mathrm{B}\approx
8.3\times10^{-14}\,\mathrm{cm^{3}\,s^{-1}}\,(T_\mathrm{II}/4\times10^4\mathrm{K})^{-1}$
is the Case-B recombination coefficient \citep{Ferland:1992aa}.
Alternatively, in the latter case, $R_\mathrm{IF}$ is approximated by
\begin{align}
r_\mathrm{neutral-flow} \equiv \left(\frac{\dot{N}_\mathrm{ion}}{4\pi
n_\mathrm{II}v_\mathrm{II}}\right)^{\frac{1}{2}} \,, 
\label{eq:r_neutral-flow}
\end{align}
where we have used the particle number conservation
$n_\mathrm{I}v_\mathrm{I}=n_\mathrm{II}v_\mathrm{II}$ across the
I-front.  The size $R_\mathrm{IF}$ is determined by the smaller between
$r_\mathrm{Strm}$ and $r_\mathrm{neutral-flow}$.  For the cases explored
in our simulations, generally we have $r_\mathrm{Strm}\le
r_\mathrm{neutral-flow}$, and thus $R_\mathrm{IF} \sim r_\mathrm{Strm}$.

In the RIAF regime
($\dot{M}<L_\mathrm{Edd}/c^2$; see Eq.~\ref{eq:eta_variable}), with
Eqs.~\eqref{eq:vII_wR}, \eqref{eq:lum_from_acc}, \eqref{eq:r_strm}, and
\eqref{eq:r_neutral-flow}, we can show that the condition
$r_\mathrm{Strm} > r_\mathrm{neutral-flow}$ is satisfied if
\begin{align}\label{eq:critv}
v_\infty \gtrsim (80\,\mathrm{km/s})
\left(\frac{n_\infty\,M_\mathrm{BH}}{10^6\,M_\odot\,\mathrm{cm^{-3}}}\right)^{1/3}\,.
\end{align}
Here, we have assumed $T_\mathrm{II}=4\times10^4\,\mathrm{K}$ and
$\left<h\nu\right>\approx 41\,\mathrm{eV}$ for the power-low radiation
spectrum $L_\nu\propto \nu^{-1.5}$ with a lower cut-off at
$13.6\,\mathrm{eV}$.  Therefore, for velocities higher than the critical
velocity in Eq.~(\ref{eq:critv}), $R_\mathrm{IF}\approx
r_\mathrm{neutral-flow}$, while for velocities lower than the critical
value from Eqs.~\eqref{eq:mdot_model} and
\eqref{eq:lum_from_acc}$\,\text{--}\,$\eqref{eq:r_strm} we find:
\begin{align}
R_\mathrm{IF}\approx r_\mathrm{Strm}=\left(\frac{12\pi\eta_0\, c^4\,m_\mathrm{p}^2\,GM_\mathrm{BH}}{
\left<h\nu\right>\alpha_\mathrm{B}\,L_\mathrm{Edd}}\right)^{\frac{1}{3}}\,
r_\mathrm{BHL,II} \approx 120\, r_\mathrm{BHL,II} 
 \,,
\label{eq:R_IF2}
\end{align}
with the BHL radius for the ionized gas defined as 
\begin{align}
&r_\mathrm{BHL,II}=\frac{G M_\mathrm{BH}}{v_\mathrm{II}^2 + c_\mathrm{II}^2} 
\label{eq:BHL_II} \\ 
&\approx
\begin{cases}
\frac{G M_\mathrm{BH}}{2c_\mathrm{II}^2} 
\approx (110\,\mathrm{au})\,\left(\frac{M_\mathrm{BH}}{100\,M_\odot}\right)\left(\frac{c_\mathrm{II}}{20\,\mathrm{km/s}}\right)^{-2}
& c_\mathrm{I}< v_\infty < v_\mathrm{R}\\[6pt]
\frac{G M_\mathrm{BH}}{v_\infty^2}  
\approx (100\,\mathrm{au})\,\left(\frac{M_\mathrm{BH}}{100\,M_\odot}\right)\left(\frac{v_\infty}{30\,\mathrm{km/s}}\right)^{-2}
& v_\infty \gg v_\mathrm{R}
\end{cases} \nonumber  \,.
\end{align}
The ratio of $R_\mathrm{IF}$ to $r_\mathrm{BHL,II}$ is independent of
either $M_\mathrm{BH}$, $n_\infty$, or $v_\infty$, except for the
dependence through $T_\mathrm{II}$. Eq.~\eqref{eq:R_IF2} implies that
the BHL radius is always much smaller than the size of the I-front in
the RIAF regime.

  \subsection{Model of the shell}
\label{sec:shell-thickness}
\begin{figure}
 \centering \hspace*{-0.5cm}
 \includegraphics[width=9cm]{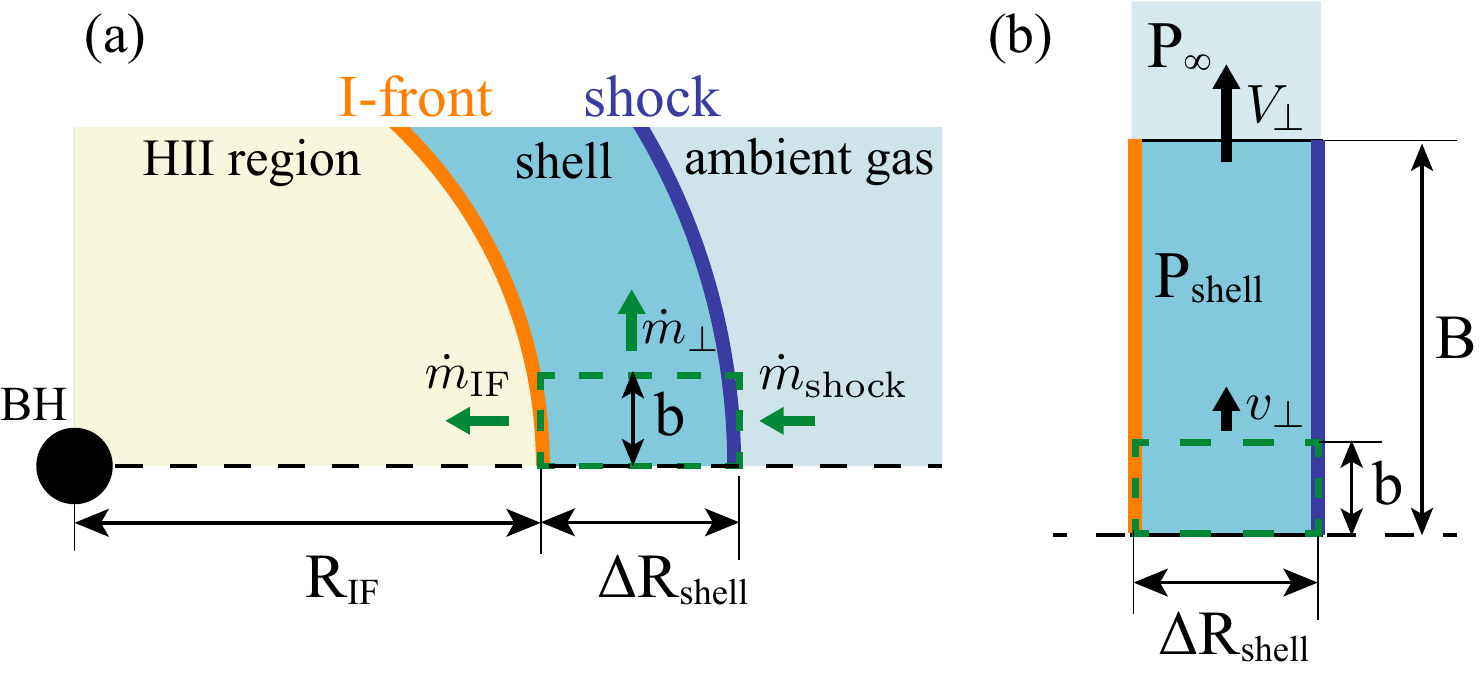} 

\caption{(a) The mass conservation in the cylindrical region of the
shell illustrated by the dashed lines. (b) Approximation of the shell
with vertical motion with a homogeneously expanding thin disk.}

 \label{fig:shell_model}
\end{figure}

In the following, we extend the Park \& Ricotti model to understand the
property of the shells forming in the D-type flows. Specifically, we
model the thickness of the shell $\Delta R_\mathrm{shell}$, which is a
critical parameter for the stability of the I-front, as we will see in
Sec.~\ref{sec:results}.  To model $\Delta R_\mathrm{shell}$, we start
from mass conservation in a cylindrical region of the shell with radius
$b$ from the axis of the BH motion (see Fig.~\ref{fig:shell_model}
a). Below, we first estimate the mass flux at each surface of the
cylinder and then derive the expression for the shell thickness.  In
this section, rather than restricting the model to isothermal shocks, we
allow the sound speed in the shell, $c_\mathrm{shell}$, to differ from
that of the incident flow $c_\mathrm{I}$. However, for simplicity, we
assume that the temperature of the shell is kept constant at $\approx
10^4\,\mathrm{K}$ due to the strong temperature dependence of Ly$\alpha$
cooling (see also Appendix~\ref{sec:X-ray-preheat}), with inefficient
cooling below $10^4$~K expected in a low-metallicity gas. We also
neglect molecular hydrogen formation and cooling in the shell.

Let us start with obtaining the incoming mass flux across the shock
\begin{align}
\dot{m}_\mathrm{shock}=\rho_\infty v_\infty
\label{eq:mdot_shock}
\end{align}
and outgoing mass flux across the I-front
$\dot{m}_\mathrm{IF}=\rho_\mathrm{II} v_\mathrm{II}$.  Using
Eqs.~\eqref{eq:vII_cD} and \eqref{eq:rhoII_cD}, we rewrite the mass flux
in the HII region as
\begin{align}
\dot{m}_\mathrm{IF} = \frac{\rho_\infty (v_\infty^2+c_\mathrm{I}^2)}{2c_\mathrm{II}}\,.
\label{eq:mdot_IF}
\end{align}
Note that the expression for $\rho_\mathrm{II}$ (Eq.~\ref{eq:rhoII_cD}),
which is derived from total pressure equilibrium and conservation of
mass flux, is valid even if the temperature changes across the shock.

Next, we estimate the outgoing mass flux at the lateral surface of the
cylinder.  Here, we make a rough approximation and regard the bow-like
shell as a thin disk with radius $B\sim O(R_\mathrm{IF})$, neglecting
the curvature of the shell and the structure along the axis
(Fig.~\ref{fig:shell_model}b).  The thermal pressure of the shocked gas,
$P_\mathrm{shell}$, which is higher than that of the ambient gas,
$P_\infty$, causes the radial expansion of the disk. We estimate the gas
velocity at the outer edge as $V_\perp \sim O(v_\infty)$, considering
the approximate balance between the thermal pressure of the shell
$P_\mathrm{shell}\approx\rho_\infty v_\infty^2$ and the ram pressure of
the ambient gas $\rho_\infty V_\perp ^2$ in the rest frame of the edge.
Since the vertical velocity $v_\perp$ is proportional to $b$ in a
uniformly expanding disk, we model it as
\begin{align}
 v_\perp = \frac{b}{B}V_\perp  \equiv \alpha \frac{b}{R_\mathrm{IF}}v_\infty\,,
 \label{eq:v_vert_shell}
\end{align}
where we introduce an $O(1)$ numerical factor $\alpha$ in the last
expression to absorb the uncertainties of the model.  The density of the
shell is determined by the approximate balance between the inner thermal
pressure and outer ram pressure at the shock as
\begin{align}
 \rho_\mathrm{shell}\approx\left(\frac{v_\infty}{c_\mathrm{shell}}\right)^2\,\rho_\infty\,.
\label{eq:rho_shell}
\end{align}
Therefore, from Eqs.~\eqref{eq:v_vert_shell} and \eqref{eq:rho_shell},
the vertical mass flux $\dot{m}_\perp = \rho_\mathrm{shell}\,v_\perp$ at
radius $b$ is given by
\begin{align}
 \dot{m}_\perp = \alpha \left(\frac{b}{R_\mathrm{IF}}\right)
\left(\frac{v_\infty}{c_\mathrm{shell}}\right)^2\rho_\infty v_\infty\,.
 \label{eq:mdot_vert}
\end{align}

Finally, we obtain $\Delta R_\mathrm{shell}$ considering mass
conservation. The equation for the conservation of mass inside the
cylinder with radius $b$ and length $\Delta R_\mathrm{shell}$ is given
by
\begin{align}
 \dot{m}_\mathrm{shock} \pi b^2 -  \dot{m}_\mathrm{IF} \pi b^2
 - \dot{m}_\perp\,2\pi b \Delta R_\mathrm{shell} = 0\,,
 \label{eq:mass_cons_sh}
\end{align}
where the first, second, and third terms correspond to the flows through
the shock, I-front, and lateral surface of the cylinder, respectively.
With Eqs.~\eqref{eq:mdot_IF}, \eqref{eq:mdot_shock} and
\eqref{eq:mdot_vert}, Eq.~\eqref{eq:mass_cons_sh} can be solved for
$\Delta R_\mathrm{shell}$ as
\begin{align}
 \Delta R_\mathrm{shell} = &
\frac{1}{2\alpha}\left(\frac{v_\infty}{c_\mathrm{shell}}\right)^{-2}
\left(1-\frac{v_\infty^2+c_\mathrm{I}^2}{2c_\mathrm{II}\,v_\infty}\right) R_\mathrm{IF}
 \label{eq:sh_thickness_v1}\\
\approx &
\frac{1}{2\alpha}\left(\frac{v_\infty}{c_\mathrm{shell}}\right)^{-2}
\left(1-\frac{v_\infty}{v_\mathrm{R}}\right) R_\mathrm{IF}
\,,
 \label{eq:sh_thickness}
\end{align}
where we have made an approximation $c_\mathrm{I}\ll v_\infty$ and used
the approximate expression of the critical velocity $v_\mathrm{R}\approx
2c_\mathrm{II}$ (Eq.~\ref{eq:vR}) in the last equation. Later, we will
determine the $O(1)$ value for $\alpha$ and the effective temperature of
ionized region $T_\mathrm{II}$ by comparing
Eq.~\eqref{eq:sh_thickness_v1} to the simulation results.  Here we
define $T_\mathrm{II}$ as an effective temperature because it absorbs
the uncertainties coming from the temperature variation inside the HII
region and the simplifying assumptions in modelling
$\dot{m}_\mathrm{IF}$. This expression implies that $\Delta
R_\mathrm{shell}$ approaches zero as $v_\infty$ approaches
$v_\mathrm{R}$ from the lower side. Thus, we can consider that the
transition from the D- to R-type flow occurs at $v_\mathrm{R}$, not only
because the R-type I-fronts are allowed only for
$v_\infty>v_\mathrm{R}$, but also because the shells in the D-type flows
can have finite thickness only for $v_\infty<v_\mathrm{R}$.

Before closing this section, we shortly discuss the column density of
the shell. From Eqs.~\eqref{eq:rho_shell} and \eqref{eq:sh_thickness},
we can roughly estimate it as $m_\mathrm{p}\,\Delta N_\mathrm{shell} =
\rho_\mathrm{shell}\,\Delta R_\mathrm{shell} \sim \rho_\infty\,
R_\mathrm{IF}\, (1-v_\infty/v_\mathrm{R})$, omitting the $O(1)$
numerical factors. Using Eqs.~\eqref{eq:R_IF2} and \eqref{eq:BHL_II},
for velocities $v_\infty<v_\mathrm{R}$, we obtain the shell column
density:
\begin{align}
\Delta N_\mathrm{shell}\approx (
3.0\times10^{21}~\mathrm{cm^{-2}})
\left(1-\frac{v_\infty}{v_\mathrm{R}}\right)
\left(\frac{n_\infty\,M_\mathrm{BH}}{10^6\,M_\odot\,\mathrm{cm^{-3}}}\right)
\,,
\label{eq:DN_shell}
\end{align}
which is proportional to $n_\infty\,M_\mathrm{BH}$ with the dependence
on $v_\infty$ through the factor $(1-v_\infty/v_\mathrm{R})$. We see
that the shell is typically optically thick to ionizing radiation for
$n_\infty\,M_\mathrm{BH}> 10^3\,M_\odot\,\mathrm{cm^{-3}}$.

 \section{3D Radiation Hydrodynamics Simulations}
\label{sec:simulations}

We perform 3D radiation hydrodynamics simulations, using the same code
as in \cite{Sugimura:2017ab,Sugimura:2018aa}, which is a modified
version of a public grid-based multidimensional magnetohydrodynamics
code PLUTO 4.1 \citep{Mignone:2007aa}. In the code, a module for 1D
multi-frequency radiation transfer coupled with primordial chemical
network and thermodynamics is implemented
\citep{Kuiper:2010ab,Kuiper:2010aa,Kuiper:2011aa,Kuiper:2013aa,Hosokawa:2016aa}.
As mentioned in the introduction, 3D simulations are necessary to study
the instability of I-fronts, but they are computationally expensive and
difficult to interpret because of the complicated interplay of various
physical processes.  Therefore, we perform 3D simulations of gas
dynamics on the scale of I-front, assuming constant luminosity and
neglecting BH gravity, to study the structure and instability of I-front
in a well-controlled numerical experiment.

We adopt a spherical computational region with the BH at the origin
$r=0$ and assume that gas flows into the computational region from the
positive $x$-direction ($\theta=\pi/2$ and $\phi=0$).  We use
logarithmically spaced grids in the $r$-direction for $0.1 R_\mathrm{IF}
< r < 10 R_\mathrm{IF}$ and linearly spaced grids in the $\theta$- and
$\phi$-directions for $0<\theta<\pi$ and $0<\phi<2\pi$, respectively. To
determine the radial range, we obtain $R_\mathrm{IF}$ by performing test
runs before the science runs.

We use nested grids in each direction, to achieve sufficiently high
resolution around the part of the I-front near the axis of the BH
motion, where the instability is likely to start growing.  In the
$r$-direction, we set the base level grids with $N_r = 64$, replace the
half of the grids around $r=R_\mathrm{IF}$ with two times finer next
level grids, and repeat a similar refinement process for the highest
level grids.  Similarly, in the $\theta$- and $\phi$-directions, we set
the base level grids with $N_\theta=80$ and $N_\phi=80$, and repeatedly
replace the highest level grids around $\theta=\pi/2$ and $\phi=0$ with
finer grids.  In the typical runs, we repeat the refinement process
three times for the $r$- and $\phi$-directions and two times for the
$\theta$-direction, resulting in approximately cubic ($\Delta r \approx
r\Delta\theta \approx r\sin\theta\Delta\phi$) minimum cells with side
length $\Delta x \approx 0.01\,R_\mathrm{IF}$.  The effective resolution
of these cells is $N_r\times N_\theta \times N_\phi = 512\times
320\times 640$. For the case of D-type flow with a thin shell, we refine
the grids one more time in each direction requiring that the shell
thickness is resolved by at least 5 cells, to avoid obtaining
artificially stable results due to the lack of the resolution.

We apply flow-in outer boundary condition on the upwind side of the
hemisphere ($0<\phi<\pi/2$ and $3\,\pi/2<\phi<2\,\pi$) with the boundary
values set to those of the ambient gas, and flow-out boundary condition
on the downwind side.  For the inner boundary, we introduce a virtual
sink cell where the physical quantities are determined by the advection
through the boundary with the conservation laws.  To alleviate having a
short time step due to polar cells with small azimuthal length $r\,\sin
\theta\,\Delta \phi$, we limit the hydrodynamic calculations to the
range of $0.2<\theta < \pi-0.2$ and instead introduce virtual cells with
radial extents set by the radial grids in the polar regions.  We neglect
the BH gravity in our simulations because the BHL radius is 10 times
smaller than the inner boundary at $0.1\,R_\mathrm{IF}$ (see
Eq.~\ref{eq:R_IF2}), hence thermal and ram pressure dominate the
dynamics of gas in our computational domain.  In the liner growth
regime, BH gravity should not affect the development of the instability
at the I-front because, not only the BH gravity is weak, but also the
gravitational field is smooth on the scale of the I-front. However, the
non-linear evolution of the instability may be altered by the BH gravity
because clumps formed as a result of the front fragmentation may
gravitationally focused toward the BH. We plan to study this process
with 3D simulations including BH gravity and radiation coupled to the
accretion rate in future works.  We also neglect gas self-gravity in
this work.  Approximating the size of fragmented clumps with the shell
thickness $\Delta R_\mathrm{shell}\sim 10^3\,\mathrm{au}\,((\Delta
R_\mathrm{shell}/R_\mathrm{IF})/0.1)(M_\mathrm{BH}/10^2\,M_\odot)$
(Eqs.~\ref{eq:R_IF2} and \ref{eq:BHL_II} with
$c_\mathrm{II}=20\,\mathrm{km/s}$) and comparing it with the Jeans
length $\lambda_\mathrm{J}\equiv
\sqrt{\pi}\,c_\mathrm{s}/\sqrt{\rho\,G}\sim 3\times10^5\,\mathrm{au}\,
(T_\mathrm{I}/10^4\,\mathrm{K})^{1/2}(n_\mathrm{shell}/10^6\,\mathrm{cm^{-3}})^{-1/2}$,
we estimate that the self-gravity is negligible for BH masses
$M_\mathrm{BH}\lesssim
3\times10^4\,M_\odot\,(T_\mathrm{I}/10^4\,\mathrm{K})^{1/2}(n_\mathrm{shell}/10^6\,\mathrm{cm^{-3}})^{-1/2}$,
but the shell and its fragment may be self-gravitating for higher BH
mass.

For the initial conditions, we assume uniform density and velocity with
the ambient values and add 10 percent random density perturbations to
seed the instability.  The results, after saturation of the instability,
are insensitive to the amplitude of the initial perturbations, which we
confirm by performing additional simulations with 10 times smaller
perturbations. This also suggests that the choice of the power spectrum
of initial perturbations does not affect our main conclusions: indeed
the instability develops similarly (although less rapidly) even without
adding perturbations to the initial conditions.  However, a large-scale
gradient of the background density field may cause some effects.  A
density gradient in the direction parallel to the BH motion modulates
the incident flow density at the I-front, leading to a back and forth
motion of the I-front, and hence a modulation of the relative velocity
of the flow with respect to the I-front.  We expect that such case can
be understood by combining the results for different velocity cases.  A
density gradient in the direction perpendicular to the BH motion
produces a net angular momentum to the gas near the BH, due to the
asymmetry of angular momentum carried by accreted gas.
\cite{Sugimura:2018aa}, who studied the effect of gas angular momentum
on accretion of non-moving BHs, found that the accretion rate is
strongly suppressed if the accretion disk formed as a result of gas
angular momentum is comparable to the Bondi radius.  The effect of gas
angular momentum in the case of moving BH in a medium with a density
gradient is yet to be studied, but it will be the topic of future work.

Below, we describe the parameters examined in our simulations. In the
reference runs, we set the parameters as follows: the ambient density
$n_\infty=10^5\,\mathrm{cm^{-3}}$; the BH mass
$M_\mathrm{BH}=10^2\,M_\odot$, which is used to calculate the
luminosity; the ambient temperature $T_\infty=10^4\,\mathrm{K}$, with a
lower temperature floor at $T_\infty$ imposed for simplicity; and the BH
velocities $v_\infty/c_\infty=2$, $4$ and $7$, where $c_\infty$ is sound
speed in the neutral gas with temperature $T_\infty$.  We set the
constant luminosity to $L=2\times10^{39}\,\mathrm{erg\,s^{-1}}$, the
value predicted by the Park \& Ricotti model for
$v_\infty/c_\mathrm{I}=4$ with Eqs.~\eqref{eq:mdotD},
\eqref{eq:lum_from_acc}, and \eqref{eq:eta_variable}. We assume a
power-low spectrum $L_\nu\propto \nu^{-1.5}$ with a lower cut-off at
$13.6\,\mathrm{eV}$ \citep[e.g.,][]{Park:2011aa}.  As we will see in the
next section, the above three velocities correspond to the three
different regions of flow structures shown in Fig.~\ref{fig:mdot_PR13}:
the stable D-type flow (Region I), the unstable D-type flow (Region II),
and the R-type flow (Region III), respectively.
In addition to the reference runs, we perform the runs with various BH
velocities in the range of $1.5 \leq v_\infty/c_\infty \leq 10$ to study
the velocity dependence in detail. To investigate the dependence on the
density and BH mass, we also perform runs with
$n_\infty=10^4\,\mathrm{cm^{-3}}$ and luminosity for
a $M_\mathrm{BH}=10^2\,M_\odot$ BH
($L=2\times10^{37}\,\mathrm{erg\,s^{-1}}$), and density
$n_\infty=10^3\,\mathrm{cm^{-3}}$ with luminosity for
a $M_\mathrm{BH}=10^2\,M_\odot$ BH
($L=2\times10^{35}\,\mathrm{erg\,s^{-1}}$), and
$n_\infty=10^4$ with luminosity for a $M_\mathrm{BH}=10^3\,M_\odot$
BH ($L=2\times10^{40}\,\mathrm{erg\,s^{-1}}$). Note that all the cases examined
in this paper are in the RIAF regime where $L\propto
\dot{M}^2/M_\mathrm{BH}\propto M_\mathrm{BH}^3\,n_\infty^2$ (see
Eqs.~\ref{eq:mdot_model}, \ref{eq:lum_from_acc}, and
\ref{eq:eta_variable}).

 \section{Results}
 \label{sec:results}

 \subsection{Region I: stable D-type flows}
\label{sec:stable-D-flow}
\begin{figure}
   \centering \includegraphics[width=8cm]{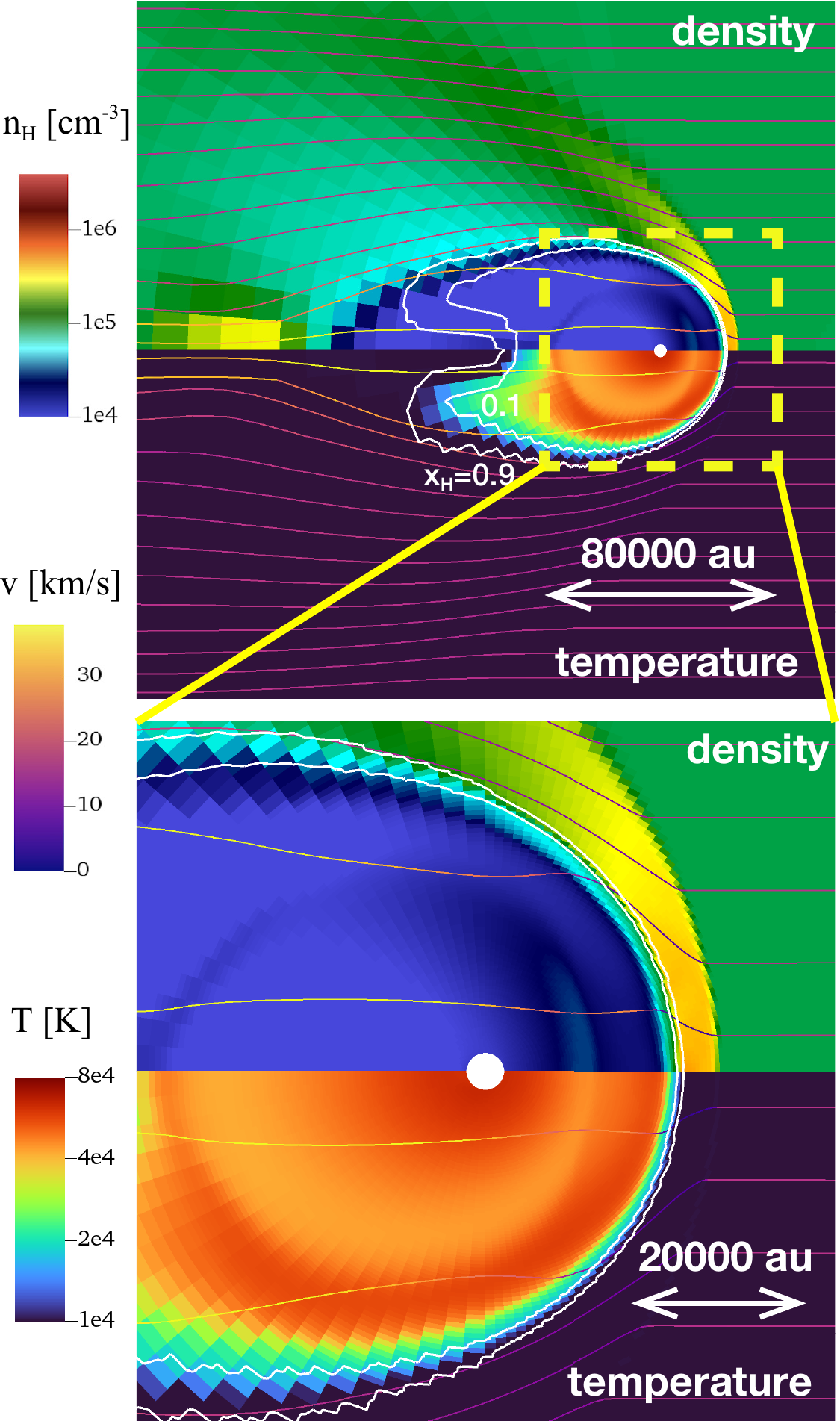}
   \caption{The snapshots of the steady-state flow in the run with
   $n_\infty= 10^5\,\mathrm{cm^{-3}}$, $M_\mathrm{BH}=10^2\,M_\odot$,
   $T_\infty=10^4\,\mathrm{K}$, and $v_\infty/c_\infty=2$. The bottom
   panel is a zoom-in view of the top panel where the entire HII bubble
   is displayed.  In each panel, we show the xy-slice of the density
   (upper) and the temperature (lower), together with the velocity
   streamlines. The outer and inner white contours correspond to the
   surfaces of the neutral fraction $x_\mathrm{HI}=0.9$ and $0.1$,
   respectively. The gas is moving from the right side to the left, with
   the BH located at the center of the sink region (white circle).  }
   \label{fig:snap_Ma2}
  \end{figure}

Figure~\ref{fig:snap_Ma2} shows the snapshot of the steady-state flow
for the run with $n_\infty= 10^5\,\mathrm{cm^{-3}}$,
$M_\mathrm{BH}=10^2\,M_\odot$, $T_\infty=10^4\,\mathrm{K}$, and
$v_\infty/c_\infty=2$.  We perform this run as the reference case for
the stable D-type flows in the Region I of
Figure~\ref{fig:mdot_PR13}. As expected, we see that a stable, dense
shell forms between the D-type I-front and the preceding shock, alike a
bow shock around a blunt body \citep[see, e.g., ][for recent studies]{Yalinewich:2016aa,Keshet:2016aa}.

Let us investigate the structure of the flow in detail.  In the HII
region, the gas is heated to the equilibrium temperature
$T_\mathrm{II}\approx 4\,\text{--}\,5\times10^4\,\mathrm{K}$, determined
by the balance of the photo-ionization heating and the Ly$\alpha$ and
free-free cooling. The shock is isothermal due to the efficient
Ly$\alpha$ cooling in the neutral gas, and the density jump in the shell
is $(v_\infty/c_\infty)^2\approx 4$ of the ambient value.  As considered
in the analytical model in Sec.~\ref{sec:analytical-model}, the gas
motion is approximately plane-parallel except for inside the shell,
where the tangentially diverging motion has a significant effect on the
streamlines.  The shell is rather thick ($\Delta
R_\mathrm{shell}/R_\mathrm{IF} \sim 0.1$) and stable.  The size of the
I-front, $R_\mathrm{IF}\sim 2\times10^4\,\mathrm{au}$, agrees with the
analytic Str\"omgren radius in Eq.~\eqref{eq:r_strm}.  In general, the
flow structure is consistent with previous 2D simulations in PR13 and
agrees with the analytical model.

To understand the properties of the shell and its stability, we
investigate the dependence of the shell thickness on the BH velocity, by
performing runs with various BH velocities
$v_\infty/c_\infty=1.5\,\text{--}\,3$ for $n_\infty=
10^5\,\mathrm{cm^{-3}}$ and $M_\mathrm{BH}=10^2\,M_\odot$. We observe
stable D-type flows for the velocity range mentioned above, but the
shell becomes unstable or disappears (the I-front becomes R-type) for velocities
$v_\infty/c_\infty>3$, as we will see in the next section.

\begin{figure*}
 \centering \includegraphics[width=14cm]{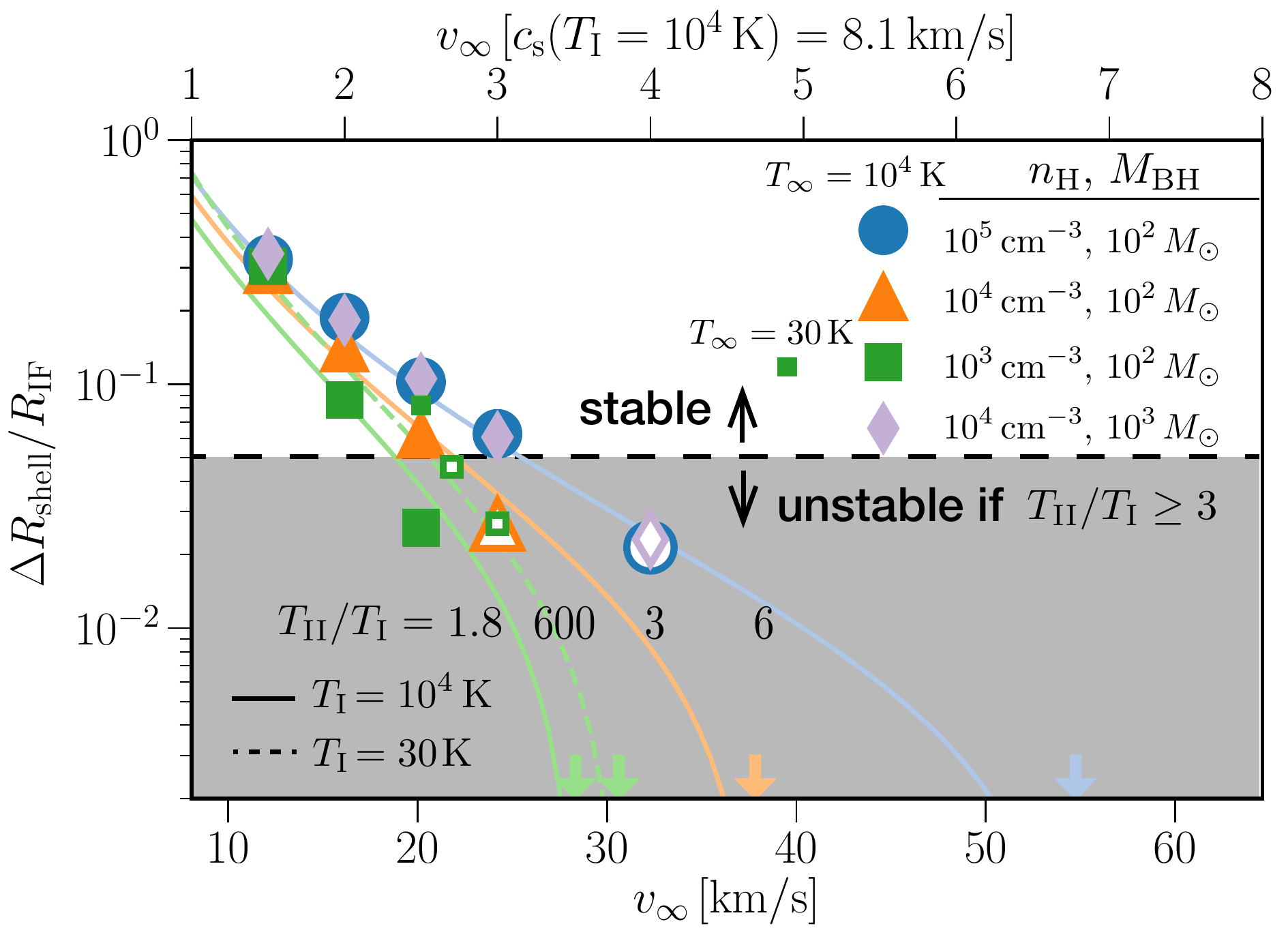}
\caption{The ratio of the shell thickness to the size of ionization
front, $\Delta R_\mathrm{shell}/R_\mathrm{IF}$, as a function of
$v_\infty$. The symbols show the results from the simulations:
$(n_\infty,\,M_\mathrm{BH})=$ $(10^5\,\mathrm{cm^{-3}},\,10^2\,M_\odot)$
(blue circles), $(10^4\,\mathrm{cm^{-3}},\,10^2\,M_\odot)$ (orange
triangles), $(10^3\,\mathrm{cm^{-3}},\,10^2\,M_\odot)$ (green squares),
and $(10^4\,\mathrm{cm^{-3}},\,10^3\,M_\odot)$ (purple diamonds). The
large symbols are for the runs with $T_\infty=10^4\,\mathrm{K}$ and the
small ones are for simulations with $T_\infty=30\,\mathrm{K}$.  The
solid lines show the analytical model in Eq.~\eqref{eq:sh_thickness_v1}
setting $\alpha=0.5$, $T_\mathrm{I}=10^4\,\mathrm{K}$ and
$T_\mathrm{II}=1.8\times10^4$ (green line), $3\times10^4$ (orange line),
and $6\times10^4$ (blue line). The green dashed line refers to a model
with $T_\mathrm{I}=30\,\mathrm{K}$ and
$T_\mathrm{shell}=10^4\,\mathrm{K}$. The lines are also labelled by the
temperature jump $\Delta_T\equiv T_\mathrm{II}/T_\mathrm{I}$ across the
I-front, that is an important parameter in predicting the stability of
the front.  The arrows indicate the value of $v_\mathrm{R}$ given by
Eq.~\eqref{eq:vR}, at which the analytical model predicts that the
thickness of the shell becomes zero.  The filled and open symbols
correspond to the stable and unstable shells, respectively.  As
discussed in the text, the front is unstable only if $\Delta
R_\mathrm{shell}/R_\mathrm{IF}<0.05$ and $\Delta_T \geq 3$, as
demarcated by the black dashed line.}  
\label{fig:ratio_shell}
\end{figure*}

Figure~\ref{fig:ratio_shell} summarizes the main results found in this
paper with regard to the stability of the I-front. The points in the
figure show the ratio of the shell thickness to the size of the I-front,
$\Delta R_\mathrm{shell}/R_\mathrm{IF}$, as a function of
$v_\infty/c_\infty$ for a large set of simulations, as shown in the
legend. We see that the ratio becomes smaller, \ie, the shell
becomes thinner, with increasing $v_\infty/c_\infty$. The filled symbols
refer to simulations in which the shell is stable, while open symbols
refer to simulations with unstable shells.

The solid lines in the figure show $\Delta
R_\mathrm{shell}/R_\mathrm{IF}$ from the analytical model described by
Eq.~\eqref{eq:sh_thickness_v1} in Sec.~\ref{sec:shell-thickness}, for
several values of $T_\mathrm{II}$, together with the arrows indicating
the values of $v_\mathrm{R}$, at which the thickness becomes zero
according to the model.  For the moment, we focus on the curve for
$T_\mathrm{II}=6\times10^4\,\mathrm{K}$ because the temperature inside
the HII region has approximately this value in the runs with
$n_\infty=10^5\,\mathrm{cm^{-3}}$ and $M_\mathrm{BH}=10^2\,M_\odot$ (see
Figure~\ref{fig:snap_Ma2}). With the numerical factor set to
$\alpha=0.5$, the analytical curve for
$T_\mathrm{II}=4\times10^4\,\mathrm{K}$ shows good agreement with the
simulation results for $n_\infty=10^5\,\mathrm{cm^{-3}}$ and
$M_\mathrm{BH}=10^2\,M_\odot$. The agreement is good also for all the
other simulations in the figure, justifying the validity of our
analytical model, as well as the choice of $\alpha=0.5$.  The analytical
model predicts that the thickness approaches zero as $v_\infty$
approaches $v_\mathrm{R}$.  In the runs with
$n_\infty=10^5\,\mathrm{cm^{-3}}$ and $M_\mathrm{BH}=10^2\,M_\odot$,
however, the shell becomes unstable before the velocity reaches
$v_\mathrm{R}$, as indicated by the open symbols.

We also investigate the dependence of shell thickness on $n_\infty$ and
$M_\mathrm{BH}$, in addition to the dependence on $v_\infty$.  We
performed runs with various $v_\infty$, assuming
$(n_\infty,\,M_\mathrm{BH})=(10^4\,\mathrm{cm^{-3}},\,10^2\,M_\odot)$,
$(10^3\,\mathrm{cm^{-3}},\,10^2\,M_\odot)$, and
$(10^3\,\mathrm{cm^{-3}},\,10^3\,M_\odot)$, and plot $\Delta
R_\mathrm{shell}/R_\mathrm{IF}$ of the stable D-type flows in
Figure~\ref{fig:ratio_shell}. We see that $\Delta
R_\mathrm{shell}/R_\mathrm{IF}$ becomes smaller when decreasing
$n_\infty$ or $M_\mathrm{BH}$. It appears that $\Delta
R_\mathrm{shell}/R_\mathrm{IF}$ is proportional to the parameter
combination $M_\mathrm{BH}\,n_\infty$.

\begin{figure}
\centering \includegraphics[width=8cm]{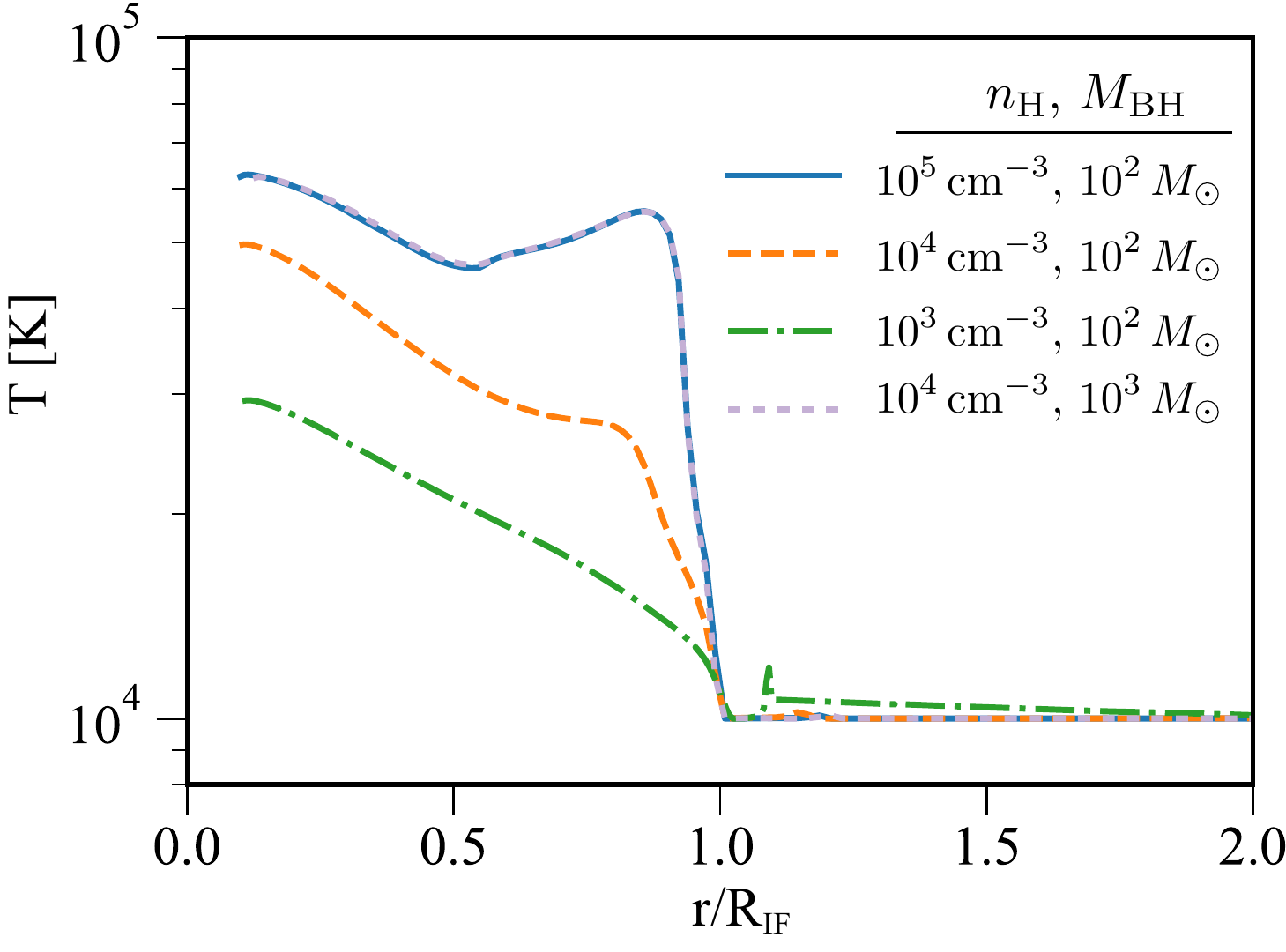}
 \caption{Upstream temperature profiles along the axis of BH motion in
 the runs with $v_\infty/c_\infty=2$ and different $n_\infty$ and
 $M_\mathrm{BH}$.  The colors represent $n_\infty$ and $M_\mathrm{BH}$
 in the same way as Fig.~\ref{fig:ratio_shell}.  The radius is
 normalized by the size of I-front, $R_\mathrm{IF}/10^4\,\mathrm{au} =
 2.2$, $1.3$, $0.92$ and $2.4$ for the runs with
 $(n_\infty,\,M_\mathrm{BH})=(10^5\,\mathrm{cm^{-3}},\,10^2\,M_\odot)$,
 $(10^4\,\mathrm{cm^{-3}},\,10^2\,M_\odot)$,
 $(10^3\,\mathrm{cm^{-3}},\,10^2\,M_\odot)$, and
 $(10^4\,\mathrm{cm^{-3}},\,10^3\,M_\odot)$, respectively.  The profiles
 for
 $(n_\infty,\,M_\mathrm{BH})=(10^5\,\mathrm{cm^{-3}},\,10^2\,M_\odot)$
 and $(10^4\,\mathrm{cm^{-3}},\,10^3\,M_\odot)$ almost completely
 overlap with each other.} \label{fig:T_profs}
\end{figure}

According to our model, the shell thickness depends on parameters other
than $v_\infty$ only because of changes of the sound speed inside the
HII region, $c_\mathrm{II}$. We will show below that $c_\mathrm{II}$
depends on $M_\mathrm{BH}\,n_\infty$, and that this dependence can be
attributed to changes in the temperature profile inside the ionized
region.  In Fig.~\ref{fig:T_profs}, we plot the upstream temperature
profiles along the axis of BH motion in the runs with
$v_\infty/c_\infty=2$ and different $n_\infty$ and $M_\mathrm{BH}$.  We
normalize the radius by the size of the I-front to directly compare the
temperature profiles.  We see in Fig.~\ref{fig:T_profs} that the
steepness of the temperature rise inside the HII region has significant
differences among the runs. For the run with BH mass
$M_\mathrm{BH}=10^2\,M_\odot$ and $n_\infty= 10^5\,\mathrm{cm^{-3}}$,
the temperature rapidly reaches the almost constant value
$5\,\text{--}\,6\times10^4\,\mathrm{K}$ inside the HII region, while the
rise in temperature becomes slower as $n_\infty$ decreases. Therefore,
in the lower-density case with $n_\infty\leq10^4\,\mathrm{cm^{-3}}$, we
need to define an effective temperature inside the HII region to be used
in the analytical model. We set the effective temperature to
$T_\mathrm{II}=1.8\times10^4\,\mathrm{K}$ and $3\times10^4\,\mathrm{K}$
for the case with $n_\infty=10^3$ and $10^4\,\mathrm{cm^{-3}}$,
respectively. These are values measured near the ionization front. In
addition, with these values adopted, the analytical curves for $\Delta
R_\mathrm{shell}/R_\mathrm{IF}$ are in good agreement with the numerical
results shown in Fig.~\ref{fig:ratio_shell}. For the case with
$n_\infty=10^3$, we also perform runs with temperature upstream of the
ionization front $T_\infty=T_\mathrm{I}=30\,\mathrm{K}$, hence assuming
that X-ray preheating is negligible (see
Appendix.~\ref{sec:X-ray-preheat} about X-ray preheating). In these
runs, we observe that the temperature rapidly increases from
$T_\mathrm{I}=30\,\mathrm{K}$ to $T_\mathrm{shell}=10^4\,\mathrm{K}$ at
the shock. The thickness of the bow-shock is slightly larger than in the
corresponding runs with $T_\infty=10^4\,\mathrm{K}$ and is well
reproduced by the analytical model with $T_\mathrm{I}=30\,\mathrm{K}$
and $T_\mathrm{shell}=10^4\,\mathrm{K}$, because the
$(1-(v_\infty^2+c_\mathrm{I}^2)/2c_\mathrm{II}v_\infty)$ term in
Eq.~\eqref{eq:sh_thickness_v1} is larger when $c_\mathrm{I}$ is lower.
As for the $M_\mathrm{BH}$ dependence, the run with
$(n_\infty,\,M_\mathrm{BH})=(10^4\,\mathrm{cm^{-3}},\,10^3\,M_\odot)$
has almost the same temperature profile as the run with
$(n_\infty,\,M_\mathrm{BH})=(10^5\,\mathrm{cm^{-3}},\,10^2\,M_\odot)$.

Next, let us explain how the temperature profile depends on $n_\infty$
and $M_\mathrm{BH}$.  The temperature just inside the I-front is roughly
determined by the balance between the photo-ionization heating and the
Ly$\alpha$ cooling.  The gas inside the I-front is almost fully ionized,
\ie, $x_\mathrm{HI}\ll 1$ and $x_\mathrm{HII}=x_\mathrm{e}\approx 1$,
where $x_\mathrm{HI}$, $x_\mathrm{HII}$, and $x_\mathrm{e}$ are the
hydrogen neutral fraction, ionized fraction, and the electron fraction,
respectively. The gas is roughly in photo-ionization equilibrium:
$k_\mathrm{ion}\,n_\mathrm{H}\,x_\mathrm{HI}=\alpha_\mathrm{B}\,n_\mathrm{H}^2\,x_\mathrm{HII}x_\mathrm{e}\approx
\alpha_\mathrm{B}\,n_\mathrm{H}^2$, where $k_\mathrm{ion}$ is the
hydrogen photo-ionization rate.  Thus, the photo-ionization heating rate
per unit volume is $\Gamma = \Delta
E\,k_\mathrm{ion}\,n_\mathrm{H}\,x_\mathrm{HI} \approx \Delta E\,
\alpha_\mathrm{B}\,n_\mathrm{H}^2$, where $\Delta E=\langle
h\nu\rangle_\mathrm{ion}-13.6\,\mathrm{eV} $ is the mean energy
deposited into the gas per photo-ionization. The Ly$\alpha$ cooling rate
can be written as $\Lambda
=\tilde{\Lambda}_\mathrm{Ly\alpha}\,n_\mathrm{H}^2\,x_\mathrm{e}\,x_\mathrm{HI}
\approx
\tilde{\Lambda}_\mathrm{Ly\alpha}\,n_\mathrm{H}^2\,x_\mathrm{HI}$ with a
$T$-dependent coefficient $\tilde{\Lambda}_\mathrm{Ly\alpha}$.  As
$\alpha_\mathrm{B}$ and $\tilde{\Lambda}_\mathrm{Ly\alpha}$ are
decreasing and increasing functions of $T$, respectively, the
equilibrium temperature (\ie, $\Gamma = \Lambda$) is higher if the
neutral fraction behind the I-front, $x_\mathrm{HI}$, is lower.

In ionization equilibrium, $x_\mathrm{HI}$ is inversely proportional to
the ionization parameter, $\xi \equiv F_\mathrm{ion}/n_\mathrm{II}$,
where $F_\mathrm{ion}$ is the flux of ionizing radiation. Since the
absorption of the ionizing photons inside the HII region is
insignificant except for the extreme vicinity of the I-front, we assume
optically thin flux in our estimate.  Then, approximating the size of
I-front with the Str\"omgren radius $r_\mathrm{Strm}$
(Eq.~\ref{eq:r_strm}), we obtain $\xi$ slightly inside the I-front as
\begin{align}
\xi = \frac{L}{4\pi\,r_\mathrm{Strm}^2\,n_\mathrm{II}}
\propto L^\frac{1}{3}\,n_\mathrm{II}^{\frac{1}{3}}\,,
\label{eq:ion_parm}
\end{align}
where we ignore the $T_\mathrm{II}$ dependence in the last expression.
From Eqs.~\eqref{eq:mdot_model}, and
\eqref{eq:lum_from_acc}-\eqref{eq:eta_variable}, the analytical model
predicts that $L$ is proportional to $M_\mathrm{BH}^3\,n_\infty^2$ in
the RIAF regime.  Therefore, from Eq.~\eqref{eq:ion_parm}, the
dependence of $\xi$ on $M_\mathrm{BH}$ and $n_\infty$ is
\begin{align}
x_{\rm HI}^{-1} \propto \xi  \propto M_\mathrm{BH}\,n_\infty\,.
\label{eq:ion_parm2}
\end{align}
This explains why the temperature rise inside the HII region is steeper
in the runs with larger $n_\infty$ for fixed $M_\mathrm{BH}$, as well as
why we obtain the same temperature profiles in runs having the same
$n_\infty\,M_\mathrm{BH}$.

\subsection{Region II: unstable D-type flows}
\label{sec:unstable-D-flow}

\begin{figure}
   \centering \includegraphics[width=8cm]{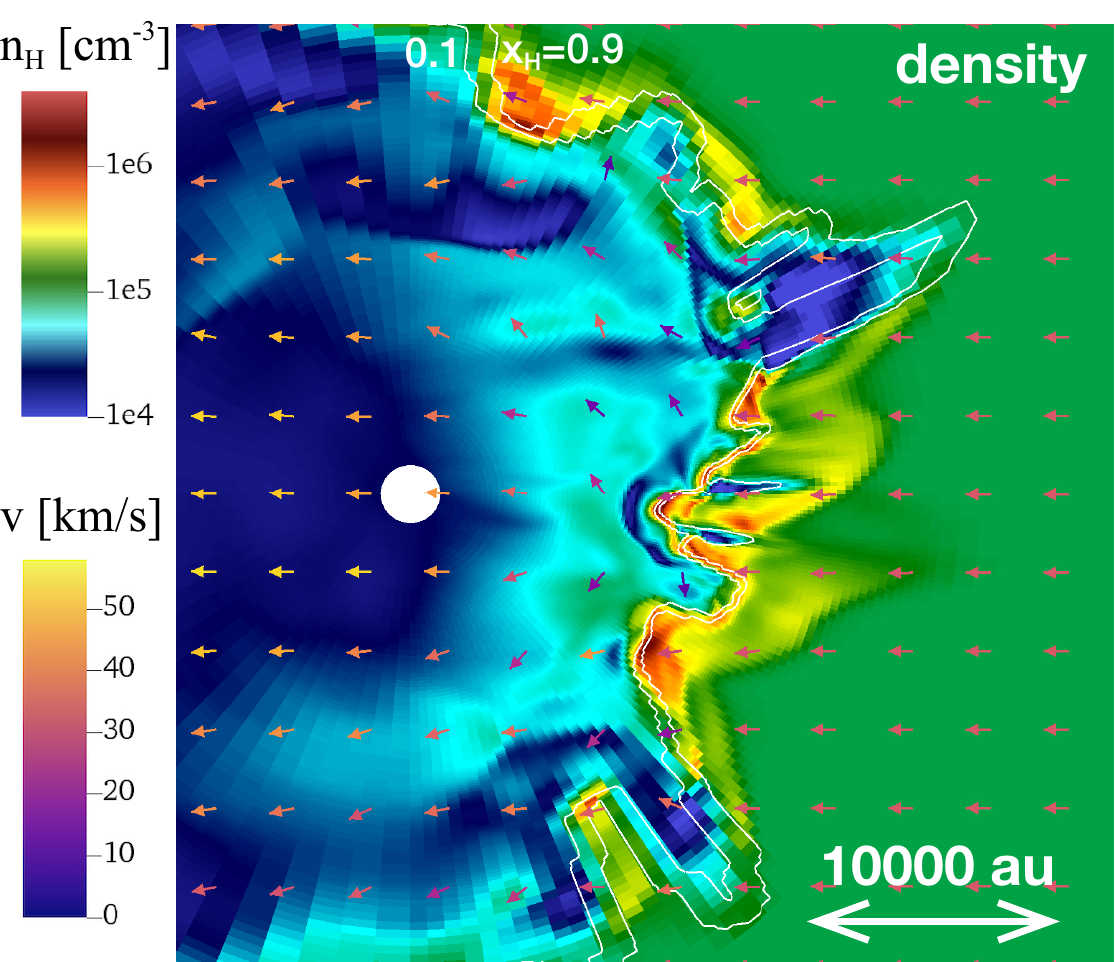}
    \caption{The same as Fig.~\ref{fig:snap_Ma2} but for the run with
    $n_\infty=10^5\,\mathrm{cm^{-3}}$, $M_\mathrm{BH}=10^2\,M_\odot$,
    $T_\infty=10^4\,\mathrm{K}$, and $v_\infty/c_\infty=4$ at a time
    after the instability is fully developed and saturated. }
    \label{fig:snap_Ma4}
  \end{figure}

\begin{figure}
   \centering \includegraphics[width=8cm]{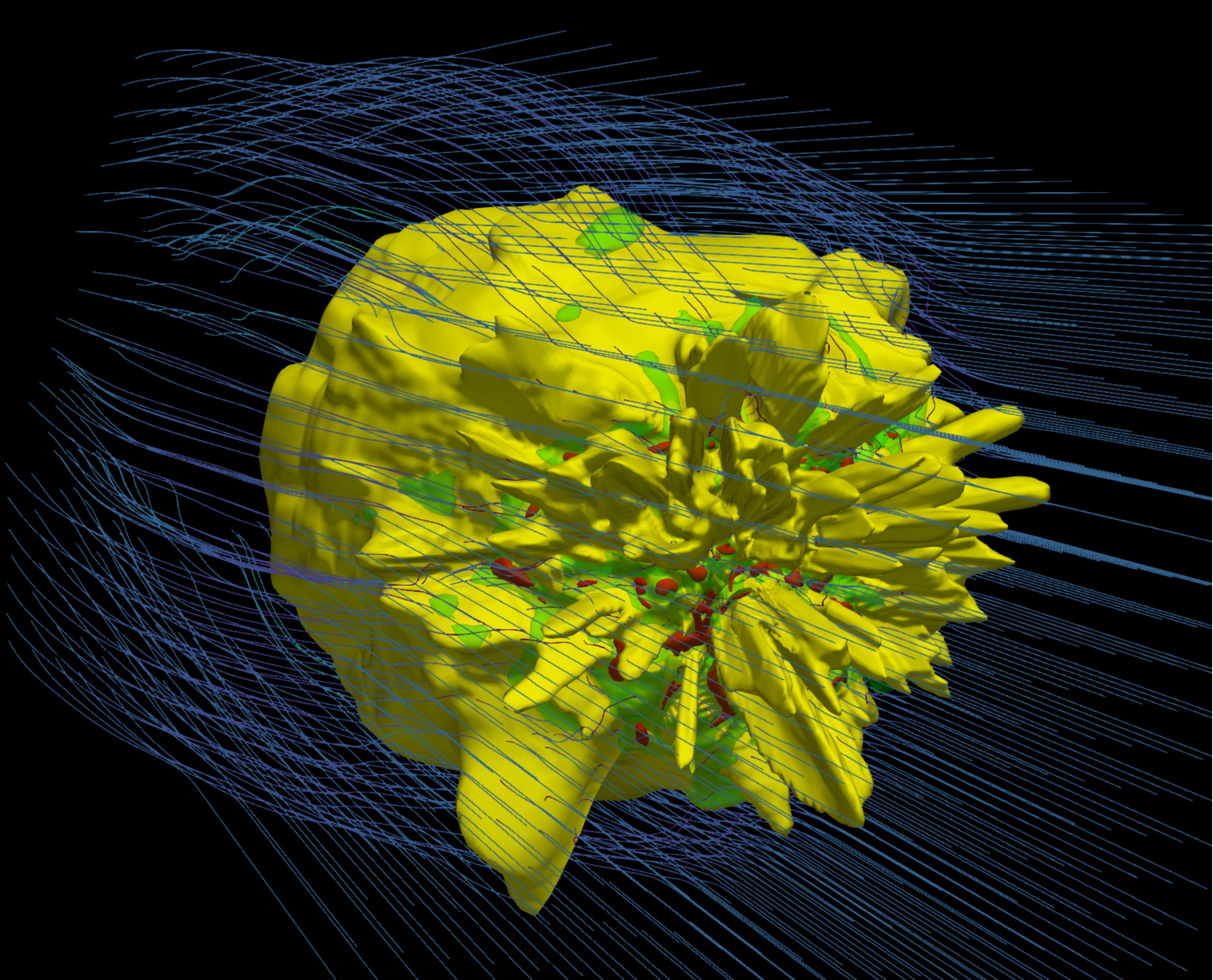}
    \caption{A 3D view of the flow structure corresponding to the 2D
    slice snapshot in Fig.~\ref{fig:snap_Ma4}.  We illustrate the
    I-front (yellow), together with the density (green and red) and
    velocity (streamlines) fields. The green and red contours correspond
    to $n_\mathrm{H}=5\times10^5$ and $10^6\,\mathrm{cm^{-3}}$,
    respectively. } \label{fig:3d_Ma4}
  \end{figure}

As we increase $v_\infty$, the dense shell in the D-type flow becomes
unstable.  As a reference case for the unstable D-type flow in the
Region II of Fig.~\ref{fig:mdot_PR13}, we perform the run with
$n_\infty= 10^5\,\mathrm{cm^{-3}}$, $M_\mathrm{BH}=10^2\,M_\odot$,
$T_\infty=10^4\,\mathrm{K}$, and $v_\infty/c_\infty=4$. We show a
snapshot of the flow structure after the instability is fully developed
and saturated in Fig.~\ref{fig:snap_Ma4}. We also provide a
corresponding 3D view in Fig.~\ref{fig:3d_Ma4}.  The shell is destroyed
by the eruptive expansions of the I-front launched from various places,
with dense clumps forming in the gaps of the I-front.

The instability in this range of $v_\infty$ was seen in the previous 2D
simulations of PR13. However, the way the instability grows is
significantly different in 2D versus 3D simulations (see Fig.8 of PR13),
as we have also confirmed by performing 2D comparative runs.  Although
the structures due to the instability are seen in various directions in
3D, they tend to converge to the axis of BH motion in 2D due to the
assumed axisymmetry, making the structures more ordered and stronger in
2D than in 3D.  Furthermore, we observe that the instability can be
artificially seeded near the axis in 2D due to the singularity of the
spherical coordinates.

The instability can be understood as the instability of a thin shell
between the D-type I-front and the shock, studied in previous
literature. This instability was found analytically using linear
analysis by \cite{Giuliani:1979aa} and confirmed in 2D
\citep{Garcia-Segura:1996aa} and 3D \citep{Whalen:2008ac,Whalen:2008aa}
simulations of an expanding D-type I-fronts around massive stars.  As
mentioned in \cite{Garcia-Segura:1996aa}, this instability is caused by
a mechanism similar to that of the thin-shell instability of an
expanding hot bubble in a cold ambient gas found by
\cite{Vishniac:1983aa}. In both cases, the thermal pressure of inner hot
gas and the ram pressure of external cold gas balanced each other in the
unperturbed state, but a force imbalance is introduced in the presence
of shell distortions because the force from the thermal pressure is
orthogonal to the surface but the ram-pressure force is parallel to the
flow, enhancing the front distortion.  Using linear perturbation
analysis, under the assumption of plane-parallel and infinitely-thin
shell, \cite{Giuliani:1979aa} showed that the modes with wavelength
$\lambda$ several times greater than the recombination length,
$d_\mathrm{r} \equiv c_\mathrm{II}/n_\mathrm{II}\alpha_\mathrm{B}$, are
unstable.

This result implies that the shell around a moving BH is unstable if the
following two conditions are satisfied.  
Firstly, the perturbed front displacement $\delta r$ needs to be larger
than the shell thickness in order for the mode to grow. Since in the
linear regime the front displacement is $\delta r \sim A \lambda$ with
$A\ll1$, we get $\lambda > \Delta R_\mathrm{shell}/A$. Since the longest
wavelengths growing modes have $\lambda< R_\mathrm{IF}$, we find that
unstable modes exist only in the range $\Delta R_\mathrm{shell}/A <
\lambda < R_\mathrm{IF}$. Therefore, $\Delta R_\mathrm{IF}/A\le
R_\mathrm{IF}$, that is $\Delta R_\mathrm{IF}/R_\mathrm{IF} \le A$, is a
necessary condition for the instability.
Secondly, the condition $d_\mathrm{r} \ll R_\mathrm{IF}$ is required, in
order for some of the above modes to be unstable. In the case examined
in this paper, the second condition is safely satisfied because
$d_\mathrm{r}\sim
20\,\mathrm{au}\,(n_\mathrm{II}/10^5\,\mathrm{cm^{-3}})^{-1}\,(T_\mathrm{II}/4\times10^4\mathrm{K})^{3/2}$
is generally much smaller than $R_\mathrm{IF}$.

To understand the condition for the instability using the simulations,
we focus on the relationship between the instability and the shell
thickness. In Fig.~\ref{fig:ratio_shell} we plot $\Delta
R_\mathrm{shell}/R_\mathrm{IF}$ for the unstable simulations (open
symbols), along with the results for the stable simulations (solid
symbols) discussed in Sec.~\ref{sec:stable-D-flow}. Since the shell
thickness is not a well-defined quantity for the unstable simulations,
we plot $\Delta R_\mathrm{shell}/R_\mathrm{IF}$ obtained from
lower-resolution runs where the instability was suppressed due to lack
of resolution.  These values should be viewed with caution because the
shell thickness is resolved only with one or two cells.  In
Fig.~\ref{fig:ratio_shell}, except for the case with
$(n_\infty,\,M_\mathrm{BH},\,T_\infty)=(10^3\,\mathrm{cm^{-3}},\,10^2\,M_\odot,\,10^4\,\mathrm{K})$,
we see that the shells are unstable if
\begin{align}
\Delta R_\mathrm{shell}/R_\mathrm{IF}<0.05,
\end{align}
as demarcated by the dashed line. 
Using Eq.~\eqref{eq:sh_thickness}, we can rewrite the condition $\Delta
R_\mathrm{shell}/R_\mathrm{IF}=0.05$ into the form
$(v_\infty/c_\mathrm{I})^{-2}(1-(v_\infty/c_\mathrm{I})(c_\mathrm{I}/v_\mathrm{R}))
= 0.05$, which is a quadratic equation of $v_\infty$ for
fixed $c_\mathrm{I}=
8.1\,\mathrm{km/s}\,(T_\mathrm{I}/10^4\,\mathrm{K})^{1/2}$ and
$v_\mathrm{R}\approx 2 c_\mathrm{II}=
46\,\mathrm{km/s}\,(T_\mathrm{II}/4\times10^4\,\mathrm{K})^{1/2}$.
The positive solution of the above equation is
\begin{align}
 v_\mathrm{crit} = \frac{5}{\sqrt{2\Delta_T}}
\left(\sqrt{1+1.6\Delta_T}-1\right)c_\infty\,,
\end{align}
where we have assumed $c_\mathrm{I}=c_\infty$ and $\Delta_T\equiv
T_\mathrm{II}/T_\mathrm{I}$. Therefore the D-type I-front is unstable in
the velocity range $v_\mathrm{crit}<v_\infty<v_\mathrm{R} $. As also evident in
Fig.~\ref{fig:ratio_shell}, the velocity range in which the shell is
unstable is rather large for large temperature jump $\Delta_T$ across
the I-front, but tends to become very narrow and nearly disappear when
$\Delta_T \rightarrow 1$.  As expected, in the run shown in
Fig.~\ref{fig:snap_Ma4}, we observe the formation of an extremely thin
shell before the instability grows and destroys the shell, supporting
the hypothesis that the instability seen in this run is the thin-shell
instability found by \cite{Giuliani:1979aa}.

The instability criteria above, however, is a necessary but not a
sufficient condition, as exemplified by the run with
$n_\infty=10^3\,\mathrm{cm^{-3}}$, $M_\mathrm{BH}=10^2\,M_\odot$,
$T_\infty=10^4\,\mathrm{K}$, and $v_\infty/c_\infty=2.5$. In this run
with $\Delta_T=1.8$ (see Fig.~\ref{fig:ratio_shell}), the I-front is
stable, although the shell is thinner than the critical value $0.05$
($\Delta R_\mathrm{shell}/R_\mathrm{IF}\sim 0.02$).  We attribute the
stability of this run to the weak temperature increase across the
I-front ($\Delta_T \sim 1$), also observed in the run with the same
$n_\infty$ and $M_\mathrm{BH}$ but with $v_\infty/c_\infty=2$ in
Fig.~\ref{fig:T_profs}. In runs with the same $n_\infty$ and
$M_\mathrm{BH}$ but $\Delta_T\approx 600$ due to
$T_\infty=30\,\mathrm{K}$, the I-fronts are unstable if $\Delta
R_\mathrm{shell}/R_\mathrm{IF}\leq 0.05$ (Fig.~\ref{fig:ratio_shell}),
supporting our conjecture that $\Delta_T$ determines the strength of the
imbalance between thermal and ram pressure in the presence of shell
distortions, hence is an important parameter for the growth of the
instability. From the simulation results, we infer that $\Delta_T \geq
3$ is a necessary condition for the existence the instability.

The ionization parameter $\xi$ and $\Delta_T$ (see
Sec.~\ref{sec:stable-D-flow}) are proportional to
$L^{1/3}\,n_\mathrm{II}^{1/3}$ (Eq.~\ref{eq:ion_parm}), and hence we can
re-formulate the condition $\Delta_T \geq 3$ as
\begin{align}
L \geq (10^{37}\,\mathrm{erg/s})\,\left(\frac{n_\mathrm{II}}{10^4\,\mathrm{cm^{-3}}}\right)^{-1}\,.
\label{eq:L_crit}
\end{align}
To derive this condition, we used $L=2\times 10^{37}\,\mathrm{erg/s}$
for the runs with $n_\infty=10^4\,\mathrm{cm^{-3}}$ (see
Sec.~\ref{sec:simulations}) and $n_\mathrm{II}\approx
2\,n_\infty=2\times10^4\,\mathrm{cm^{-3}}$ with $v_\infty$ around
$v_\mathrm{R}\sim 2\,c_\mathrm{II}$ (Eq.~\ref{eq:rhoII_cD}). In the RIAF
regime, $\xi$ is also proportional to $n_\infty\,M_\mathrm{BH}$
(Eq.~\ref{eq:ion_parm2}), and thus the above condition is equivalent to
\begin{align}
n_\infty\,M_\mathrm{BH} \geq 10^6\,M_\odot\,\mathrm{cm^{-3}}\,.
\end{align}
In addition, using Eqs.~\eqref{eq:rhoII_cD}
and \eqref{eq:R_IF2}, we find $d_\mathrm{r}/R_\mathrm{IF} \propto
(M_\mathrm{BH} n_\infty)^{-1}$. Hence, the condition for the existence
of unstable modes, $d_\mathrm{r}/R_\mathrm{IF}< 1$, is more strongly
satisfied with increasing $n_\infty\,M_\mathrm{BH}$.  Although we have
confirmed that  $d_\mathrm{r}/R_\mathrm{IF} < 1$ in our simulations
with  $n_\infty\,M_\mathrm{BH} \geq
10^5\,M_\odot\,\mathrm{cm^{-3}}$, the relatively larger
$d_\mathrm{r}/R_\mathrm{IF}$ in the run with
$n_\infty=10^3\,\mathrm{cm^{-3}}$, $M_\mathrm{BH}=10^2\,M_\odot$ and
$v_\infty/c_\infty=2.5$ may contribute to suppressing the instability.
Finally, we also note that the opacity to ionizing radiation of shell is
also proportional to $n_\infty\,M_\mathrm{BH}$, as shown in
Eq.~\eqref{eq:DN_shell}. Thus, the lower opacity and self-shielding of
the shell in the cases with smaller $n_\infty\,M_\mathrm{BH}$ may also
have a stabilizing effect.  In summary, the parameter combination
$n_\infty\,M_\mathrm{BH}$ controls the nature of the flow and the
stability of I-front around moving BHs. As inferred from our simulation
results, we regard the condition $n_\infty\,M_\mathrm{BH} \geq
10^6\,M_\odot\,\mathrm{cm^{-3}}$ as the second necessary condition for
having unstable and fragmenting shells around moving BHs.

\subsection{Region III: unstable R-type flows}
\label{sec:unstable-R-flow}
\begin{figure}
   \centering \includegraphics[width=8cm]{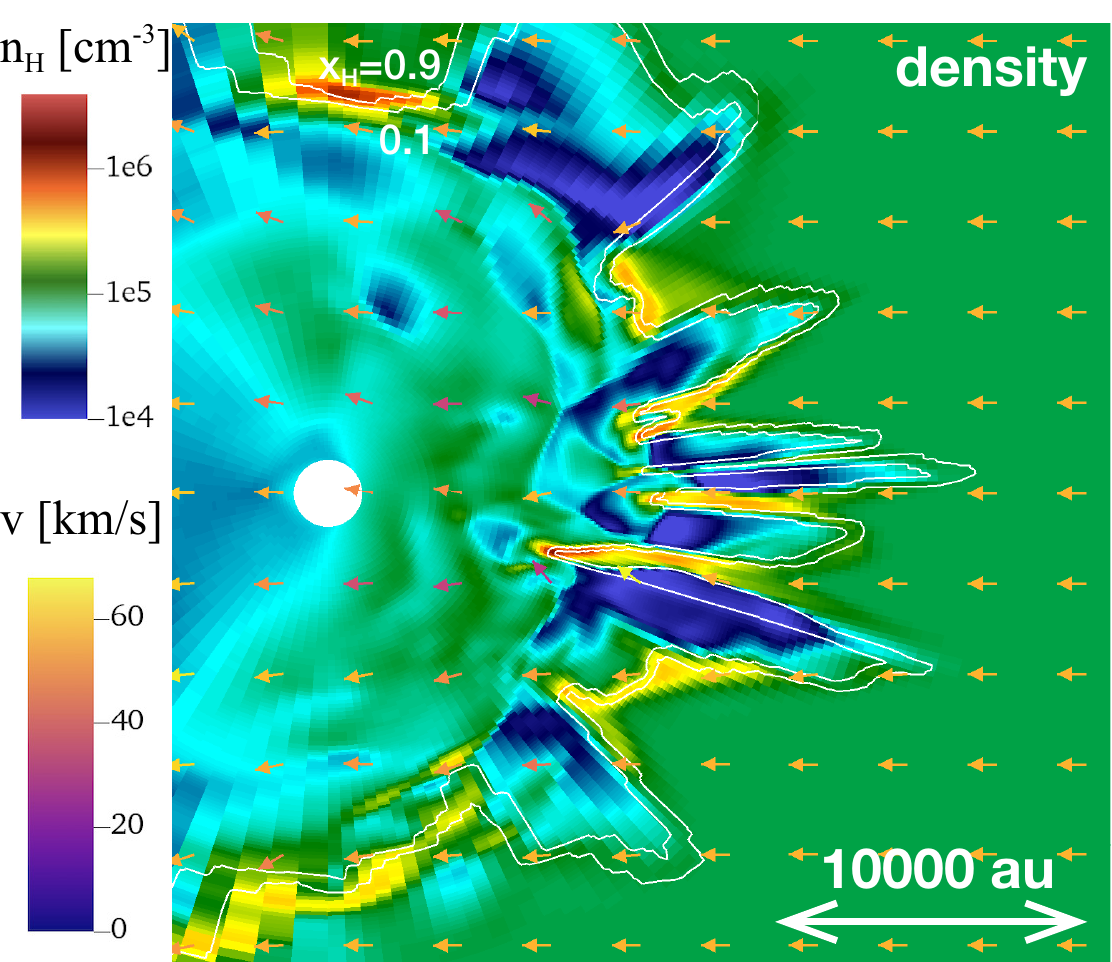}
   \caption{The same as Fig.~\ref{fig:snap_Ma2} but for the run with
   $n_\infty= 10^5\,\mathrm{cm^{-3}}$ and $v_\infty/c_\infty=7$ at a
   time after the instability is fully developed and saturated.}
   \label{fig:snap_Ma7}
  \end{figure}

For velocities $v_\infty \gtrsim v_{\rm R}$, the dense shell disappears
and the flow becomes R-type.  For a range of velocities around the
critical value, however, the transition between the D- and R-type flows
is difficult to define because the I-front is not steady due to the
instability.  As a reference case for the R-type flow in the Region III
in Figure~\ref{fig:mdot_PR13}, we perform the run with $n_\infty=
10^5\,\mathrm{cm^{-3}}$, $M_\mathrm{BH}=10^2\,M_\odot$,
$T_\infty=10^4\,\mathrm{K}$, and $v_\infty/c_\infty=7$ and shows the
snapshot at a time after the instability is fully developed and
saturated in Figure~\ref{fig:snap_Ma7}.  Unexpectedly, with respect to
the results in PR13, we find that the R-type I-front is also
unstable. Similarly to the case of the unstable D-type flow, in the
non-linear regime, the instability strongly distorts the I-front and
dense regions form at the receding gaps of the I-front.  This type of
instability was not observed in PR13, probably because the resolution at
the I-front was not sufficiently high due to the logarithmic grid in the
radial direction.  By performing low-resolution simulations, we find
that the instability disappears when the resolution near the I-front is
below $\Delta x/R_\mathrm{IF} = 0.05 $.  We would also like to emphasize
again that the growth of instability is essentially a 3D process and
cannot be fully captured by 2D simulations.

The unstable nature of weak R-type I-front has been known since the late
1960s.  \cite{Newman:1967aa} studied the stability of the weak R-type
I-front using linear analysis for a plane-parallel and infinitely-thin
I-front. \cite{Newman:1967aa} explained the mechanism of instability as
follows: the parts of the perturbed I-front getting displaced upstream,
the relative flow velocity with respect to the I-front becomes higher;
thus the density behind the I-front becomes smaller according to the
jump condition for weak R-type I-front (see Eq.~\ref{eq:jump_IF} with
minus sign). Consequently, the absorption of ionizing photons by
recombining neutral hydrogen is suppressed, leading to a further
displacement upstream of the I-front. Vice versa for the perturbed
I-front displaced in the downstream direction. The linear analysis
showed that all modes are unstable irrespective of their wavelength
$\lambda$, and the growth timescale becomes shorter at smaller
wavelengths, until $\lambda \simeq d_\mathrm{r}$, below which the growth
time scale remains nearly constant. In less idealized conditions,
however, we expect that the instability for small-scale modes may be
suppressed by effects not included in the linear analysis, such as the
finite thickness of the I-front and possible presence of magnetic fields
\citep{Axford:1964aa}.  For a mode with fixed $\lambda$, the growth time
scale of the instability is nearly constant with the velocity
$v_\mathrm{I}$, but decreases at most by a factor of two as
$v_\mathrm{I}$ approaches $v_\mathrm{R}$.  The growth time scale of the
fastest modes is of the order of the recombination time scale
$t_\mathrm{rec}=1/n_\mathrm{II}\alpha_\mathrm{B}$.

While \cite{Newman:1967aa} showed that the weak R-type I-fronts are
always unstable, they concluded that the growth of the instability was
insignificant because the duration of the expanding R-type phase of an
HII region is only about $t_\mathrm{rec}$, and thus the instability does
not have enough time to grow substantially.  In the current case of
moving BH, however, weak R-type I-fronts continue to exist for a much
longer time, and thus there is enough time for the instability to grow
until it becomes non-linear and saturates.

As shown in Fig.~\ref{fig:snap_Ma7}, the simulations results confirm the
linear analysis expectations: in the runs with $n_\infty=
10^5\,\mathrm{cm^{-3}}$, $M_\mathrm{BH}=10^2\,M_\odot$, and
$v_\infty/c_\infty=7$ and $10$, the I-front is unstable. At lower
ambient density, in the runs with $n_\infty=10^4\,\mathrm{cm^{-3}}$,
$M_\mathrm{BH}=10^2\,M_\odot$, and $v_\infty/c_\infty=4\,\text{--}\,10$,
we again observe the instability of the R-type flows. The distortion of
I-front and the density contrast of the front fragments, however,
becomes less significant than in the higher density runs. With an even
weaker temperature jump, in the runs with
$n_\infty=10^3\,\mathrm{cm^{-3}}$, $M_\mathrm{BH}=10^2\,M_\odot$, and
$v_\infty/c_\infty=3\,\text{--}\,10$, the instability of the R-type
flows becomes insignificant. We further confirm the requirement of large
$\Delta_T$ for the growth of I-front instabilities, by performing runs
with $\Delta_T\approx 600$, assuming
$T_\mathrm{I}=T_\infty=30\,\mathrm{K}$ (X-ray preheating artificially
suppressed). We find that the R-type I-front is unstable in the run with
$v_\infty=4\,c_\mathrm{s}(T=10^4\,\mathrm{K})=70\,c_\infty$.  The
instability, however, does not grow in the run with
$v_\infty=7\,c_\mathrm{s}(T=10^4\,\mathrm{K})=130\,c_\infty$.  This case
suggests that the instability may become weak if the Mach number is very
large, although this case is unlikely to be realized in nature because
we expect that $T_\mathrm{I}\sim 10^4\,\mathrm{K}$ due to X-ray
preheating, especially for R-type flows in which a shell optically thick
to X-rays is not present. In the runs with
$n_\infty=10^4\,\mathrm{cm^{-3}}$, $M_\mathrm{BH}=10^3\,M_\odot$, and
$v_\infty/c_\infty=7$ and $10$, the front is strongly unstable similarly
as in the reference run, because the temperature profiles are identical
in these two cases that have the same $n_\infty\,M_\mathrm{BH}$.

In the same way as the thin-shell instability of the D-type flow, the
steep temperature jump is a necessary condition in order for the
instability to grow.  As mentioned above, the instability is driven by
the variation of the density behind the I-front as the flow velocity
varies behind the perturbed front.  The density behind the front depends
on the flow velocity more weakly as the ratio $\Delta_T$ approaches
unity (Eq.~\ref{eq:jump_IF}).  Therefore, a steeper temperature jump
enhances the strength of the instability.  The condition for a steep
enough temperature jump is to have a large ionization parameter $\xi$
just behind the I-front.  In our constant luminosity simulations of the
D-type and R-type flows, we assume a constant luminosity $L$ at the
value $v_\infty=4\,c_\infty$, irrespective of the actual value of
$v_\infty$ in the runs (see Sec.~\ref{sec:simulations}). This is a good
approximation for unstable D-type flows in the range of
$v_\mathrm{crit}<v_\infty<v_\mathrm{R}$, as well as for R-type flows
with $v_\infty \gtrsim v_\mathrm{R}$, because for this velocity range
the luminosity does not change significantly as a function of
$v_\infty$. With this assumption we have found that the I-front is
unstable owing to large enough $\xi$ if $n_\infty\,M_\mathrm{BH} \geq
10^6\,M_\odot\,\mathrm{cm^{-3}}$.

However, for the R-type flows with large $v_\infty$ (\ie, when $v_\infty
\gg v_\mathrm{R}$), the reduction of $L$ from the value assumed in our
constant luminosity simulations can be significant.  Therefore, here we
derive a more accurate instability criterion allowing a realistic
dependence of $L$ on $v_\infty$. In the high-velocity limit
($v_\infty\gg v_\mathrm{R}$), Eqs.~\eqref{eq:mdot_model} and
\eqref{eq:mdotR}-\eqref{eq:eta_variable}, give $L\propto
M_\mathrm{BH}^3\,n_\infty^2v_\infty^{-6}$, and thus
Eq.~\eqref{eq:ion_parm} yields $ \xi \propto
L^{1/3}\,n_\mathrm{II}^{1/3} \propto
M_\mathrm{BH}\,n_\infty\,v_\infty^{-2}$.  From
Eqs.~\eqref{eq:mdot_model}$\,\text{--}\,$\eqref{eq:eta_variable}, the
actual value of $L$ for $v_\infty=100\,\mathrm{km/s}$ is about $10^3$
times smaller than the assumed constant value of $L$ for
$v_\infty=4\,c_\infty$. Therefore, by rewriting the condition of $\xi$
larger than the critical value corresponding to $n_\infty\,M_\mathrm{BH}
= 10^6\,M_\odot\,\mathrm{cm^{-3}}$ and $L$ for $v_\infty=4\,c_\infty$
with Eq.~\eqref{eq:ion_parm}, we obtain the instability condition for
R-type I-fronts:
\begin{align}
n_\infty\,M_\mathrm{BH} \geq (10^7\, M_\odot\,\mathrm{cm^{-3}})\,
\left(\frac{v_\infty}{100\,\mathrm{km/s}}\right)^2 \qquad (v_\infty\gg v_\mathrm{R})\,.
\end{align}
We have confirmed the validity of this condition by performing
additional runs with $n_\infty=10^4\,\mathrm{cm^{-3}}$,
$v_\infty=13\,c_\infty\approx 100\,\mathrm{km/s}$, and two different
masses: $M_\mathrm{BH}=10^2$ and $10^3\,M_\odot$, for which we determine
$L$ including the $v_\infty$ dependence.  The I-front in the run with
the smaller mass is stable, while in the other run is unstable,
confirming the validity of our stability criteria.

\section{Discussions}
\label{sec:discussion}

\subsection{Conditions for Instability}
In Sec.~\ref{sec:results}, we have observed that I-fronts around moving
BHs can be either stable or unstable, depending on the parameters of the
system.  Below, we will summarize and discuss the conditions for the
instability based on our results.

We have found that a sufficiently large temperature jump, $\Delta_T >3$,
across the I-front is a necessary condition for the instability.  This
first condition is satisfied if $L \geq
10^{37}\,\mathrm{erg/s}\,(n_\mathrm{II}/10^4\,\mathrm{cm^{-3}})^{-1}$,
which is equivalent to $n_\infty\,M_\mathrm{BH} \geq
10^6\,M_\odot\,\mathrm{cm^{-3}}$ for D-type flows (and low-velocity
R-type flows) and $n_\infty\,M_\mathrm{BH} \geq 10^7\,M_\odot\,\mathrm{cm^{-3}}\,
(v_\infty/100\,\mathrm{km/s})^2$ for
high-velocity R-type flows (with $v_\infty\gg v_\mathrm{R}$). This
condition is also satisfied if $T_\mathrm{I}\lesssim5000\,\mathrm{K}$
(given that $T_\mathrm{II}$ has a lower floor $\sim 1.5\times
10^4\,\mathrm{K}$ for zero-metallicity gas). However, even if the
ambient temperature $T_\infty$ is cold, the incident flow is heated to a
temperature $T_\mathrm{I}\sim 10^4\,\mathrm{K}$ by X-ray pre-heating
before reaching either the shock (for D-type flows) or the I-front (for
R-type flows), unless the dense shell preceding D-type fronts is
optically thick to X-rays (see Appendix \ref{sec:X-ray-preheat}).

In order for the instability to exist, the flow needs to be either
D-type with a sufficiently thin shell ($\Delta
R_\mathrm{shell}/R_\mathrm{IF}< 0.05$) or an R-type flow. This
requirement yields the second condition for the instability:
$v_\infty/c_\infty \geq 3\,\text{--}\,4$, or $v_\infty \geq
20\,\text{--}\,30\,\mathrm{km/s}$, with the minimum value slightly
depending on the temperature profile in the HII region.

Now, let us discuss the (in)stability of the I-fronts around actual
moving BHs, by applying the above two conditions to astrophysical
situations.  The first condition of $n_\infty\,M_\mathrm{BH}$ is rarely
satisfied if we consider stellar-mass BHs with $M_\mathrm{BH} <
10^2\,M_\odot$ moving in local galaxies because $n_\infty\geq
10^4\,\mathrm{cm^{-3}}$ required for the instability is significantly
higher than the mean density of typical molecular clouds ($\sim
10^2\,\mathrm{cm^{-3}}$). However, this does not exclude the occurrence
of the instability. IMBHs with $M_\mathrm{BH} =
10^{5\,\text{--}\,6}\,M_\odot$ are expected to exhibit the front
instability even as moving in the cold and warm atomic phases of the
interstellar medium $\sim 1\,\text{--}\,10\,\mathrm{cm^{-3}}$.
Moreover, the instability is expected to be important in high-redshift
galaxies as their interstellar medium and molecular clouds have mean
densities higher than that of local galaxies.

The second condition of $v_\infty$ is easily satisfied.  If we consider
BHs moving at about the circular velocity of the host galaxies ($\sim100
\mathrm{km/s}$), the velocity of such BHs is larger than the value
needed for the instability ($20\,\text{--}\,30\,\mathrm{km/s}$), as well
as the critical velocity for the D- and R-type transition
($v_\mathrm{R}\sim 30\,\text{--}\,55\,\mathrm{km/s}$ depending on the
temperature profile).  Therefore, the R-type flows around these BHs are
unstable if the first condition of $n_\infty\,M_\mathrm{BH}$ is
satisfied. If the dynamical friction by stars and gas works effectively
and reduces the relative velocity between the gas and the BHs, the flows
first become D-type with possibly-unstable thin shells and then D-type
with stable thick shells.  The effect of the gas dynamical friction,
however, can be weakened or even reversed due to the gravitational
attraction by the shells of the D-type flows
\citep{Park:2017aa,Toyouchi:2020aa}.

\subsection{Accretion and Radiation Bursts}
The instability of the I-fronts observed in our simulations may lead to
the accretion and radiation bursts.  The dense clumps forming as a
result of the instability possibly fall directly into the BHL radius and
cause accretion bursts (see PR13). The associated radiation bursts would
affect the gas dynamics on the scale of the I-front, producing periodic
break up the I-front.  The interaction of the gas dynamics and the
time-varying radiation flux produces a complicated feedback loop, which
will be addressed in future work.

The accretion and radiation bursts can have various astrophysical
consequences.  The burst-mode accretion may enhance the time-averaged
accretion rate and alleviate the problem of the inefficient seed IMBH
growth in models of SMBH formation \citep[see,
e.g.,][]{Sugimura:2018aa}.  Even if the time-averaged accretion rate is
not modified, the accretion bursts can increase the total amount of
X-ray emission because the radiative efficiency is higher for a higher
accretion rate in the RIAF regime.  The modified X-ray emission can
affect the star-formation history of the Universe through the build-up
of an X-ray background \citep[see, e.g.,][]{Ricotti:2016ab}.  Finally,
the radiation bursts may temporarily enhance the luminosity of moving
IMBHs not detectable without bursts to the range of the ultraluminous
X-ray sources \citep[ULXs; see, e.g.,][for review]{Miller:2004aa},
opening the way to the observation of such IMBHs in nearby galaxies.
We will discuss in more detail this last point in the next section.

\subsection{X-rays from Moving IMBHs: Observational Implications}

Although more work needs to be done to correctly link the I-front
instability to the change of BH accretion rate and luminosity, here we
briefly discuss possible observational signatures and tests of our
model. We first discuss observations of steady radiation from moving BHs
and then make qualitative prediction considering the possibility of
luminosity bursts associated with the I-front instability.

Currently, observations have sufficient sensitivity to detect X-ray (and
possibly radio) emission from moving IMBHs accreting from molecular
clouds in the interstellar medium
\citep[e.g.,][]{Ipser:1977aa,Fujita:1998aa,Gaggero:2017aa,Ioka:2017aa,Matsumoto:2018aa,Manshanden:2019aa}.
There are two promising sites for finding such IMBHs. The first is the
Galactic center, where the gas density is high and IMBHs may be more
concentrated within the Galaxy.  At the Galactic center, the steady
accretion model in Sec.~\ref{sec:1d-model} and
Appendix~\ref{sec:analytical-model-appendix} predicts that IMBHs with
$M_\mathrm{BH}\gtrsim 10^2\,M_\odot
(n_\mathrm{H}/10^2\,\mathrm{cm^{-3}})^{-2/3}$ have X-ray luminosity
$L_\mathrm{X}>4\times10^{32}\,\mathrm{erg/s}$, which is the luminosity
limit of the {\it Chandra} X-ray source catalog \citep{Muno:2009aa}. For
this rough estimate we have assumed that IMBH velocity is $\sim
40\,\mathrm{km/s}$, hence the luminosity is near the peak value, and
that the X-ray luminosity is $30\,\%$ of the bolometric luminosity
\citep{Manshanden:2019aa}.  The second strategy to find IMBHs is in
dwarf galaxies, because they are gas rich and because of the larger
number of IMBHs with velocities near 40 km/s (due to the lower virial
velocity of dwarf galaxies with respect to the Milky Way).  {\it
Chandra} provided X-ray source catalog for sources with
$L_\mathrm{X}>4\times
10^{35}\,\mathrm{erg/s}\,(d_\mathrm{L}/2\,\mathrm{Mpc})^2$ in the COSMOS
field \citep{Civano:2016aa}, and thus IMBHs with masses
$M_\mathrm{BH}\gtrsim
10^3\,M_\odot\,(n_\mathrm{H}/10^2\,\mathrm{cm^{-3}})^{-2/3}(d_\mathrm{L}/2\,\mathrm{Mpc})^{2/3}$
are detectable if we make the same assumptions as above.

The luminosity bursts associated with the I-front instability are
expected to be significantly brighter than the steady state luminosity,
but short lived. Therefore, accounting for these luminosity bursts,
there will be a small fraction (due to the short duty cycle of the
burst) of the accreting IMBH population with higher luminosity.  As a
result, the X-ray luminosity function may show a small (in term of
number of object) high-luminosity component above the critical
luminosity for the instability, $\sim
10^{37}\,\mathrm{erg/s}\,(n_\mathrm{II}/10^4\,\mathrm{cm^{-3}})$
(Eq.~\ref{eq:L_crit}).  While such luminous objects have already been
strongly constrained near the Galactic center, the precision of the
luminosity function in dwarf galaxies will be greatly improved in the
future. For example, {\it Athena} plans an X-ray survey that increases
the area by one order of magnitude and sensitivity by two orders of
magnitude with respect to the current observational limits
\citep{Aird:2013aa}.  Detailed modeling of the X-ray luminosity function
in the Galactic center and nearby dwarfs may provide a test of our model
for the instability of the I-front.  However, in order to estimate the
increase of luminosity during the bursts due to fragment of the shell
falling into the IMBH, we need full 3D simulations of moving BHs with
gravity and radiative feedback.

 \section{Summary}
\label{sec:summary}

In this paper, we have studied the structure and instability of the
ionization fronts (I-fronts) around moving black holes (BHs), by
performing a series of 3D radiation hydrodynamics simulations with
high-resolution around the I-front, and assuming a constant luminosity
from the BH.  To provide a physical interpretation of the structures
observed in the simulations, we have developed an analytical model of
the bow-shock preceding the D-type I-fronts and studied the conditions
for their instability.

The I-front makes a transition from D- to R-type as we increase the BH
velocity $v_\infty$, and the D-type front becomes unstable for
velocities above a critical value $v_\infty \approx
20\,\text{--}\,30\,\mathrm{km/s}$, when the ratio of the shell thickness
to the size of the I-front is smaller than $0.05$. The R-type flow is
instead always unstable irrespective of $v_\infty$, with the possible
exception of very large Mach numbers ($> 100$). However, the instability
grows only if $n_\mathrm{H}\,M_\mathrm{BH} \gtrsim
10^6\,M_\odot\,\mathrm{cm^{-3}}$, for which the ionization parameter at
the I-front and therefore the temperature jump $\Delta_T\equiv
T_\mathrm{II}/T_\mathrm{I}$ is larger than a critical value ($\Delta_T
\ge 3$).  Although this last condition is rarely satisfied for
stellar-mass BHs moving in local molecular clouds, we expect that
intermediate-mass BHs (IMBHs) moving in the denser interstellar medium
(ISM) of galaxies in the early Universe would exhibit I-front
instabilities, which may lead to accretion and radiation bursts. Such
bursts would increase the IMBH growth rate, contribute to the build-up
of the X-ray background, and improve the observability of IMBHs
accreting from molecular clouds.

This paper is a first step to better understand the entire process of
gas accretion onto moving IMBHs, including the possible accretion and
radiation bursts due to the instability of the I-front.

\section*{Acknowledgements}
The authors thank Takashi Hosokawa, Daisuke Toyouchi, and Hidenobu
Yajima for fruitful discussions and comments.  K.S. appreciates the
support by the Fellowship of the Japan Society for the Promotion of
Science for Research Abroad.  M.R. acknowledges the support by NASA
grant 80NSSC18K0527. The numerical simulations were performed on the
Cray XC50 at CfCA of the National Astronomical Observatory of Japan, as
well as on the computer cluster, Draco, at Frontier Research Institute
for Interdisciplinary Sciences of Tohoku University, and on the Cray
XC40 at Yukawa Institute for Theoretical Physics in Kyoto
University. The authors also acknowledge the University of Maryland
supercomputing resources (http://hpcc.umd.edu) made available for
conducting the research reported in this paper. We thank the
anonymous reviewer for helping improve the quality of this paper.




   \bibliographystyle{mnras}


\appendix
  \section{Revised accretion rate formula including X-ray preheating
 and extension to subsonic velocities}
\label{sec:analytical-model-appendix}

Here, we provide an accretion rate formula that includes X-ray
preheating and extension to subsonic velocities, by revising the Park \&
Ricotti model described in Sec.~\ref{sec:analytical-model}.

  \subsection{Subsonic case}
\label{sec:subsonic-case}

Let us begin by describing the subsonic case with $v_\infty <
c_\mathrm{I}$.  In this case, not only is the shocked shell of the
D-type flow replaced with a gradual density enhancement, but also the
effect of the vertical motion relative to the parallel motion becomes
significant and the approximation of the 1D plane-parallel flow breaks
down as the velocity decreases.  Therefore, we interpolate the accretion
rates at $v_\infty= c_\mathrm{I}$ and $0$, instead of modelling the
accretion rate with the 1D plane-parallel model.

In the limit of $v_\infty=0$, the system becomes the well-known
spherical symmetric case, in which the accretion is episodic, but the
average accretion rate is approximately given by the Bondi accretion
rate for the ionized region
\citep{Milosavljevic:2009ab,Park:2011aa}. With the assumption of
pressure equilibrium between the inner ionized and outer neutral gas,
$\rho_\infty c_\mathrm{I}^2=\rho_\mathrm{II} c_\mathrm{II}^2$, the
accretion rate is given by $\dot{M}_\mathrm{v=0}=4\pi \lambda_\mathrm{B}
G^2 M_\mathrm{BH}^2\rho_\infty c_\mathrm{I}^2/c_\mathrm{II}^5 $, where
we set the coefficient $\lambda_\mathrm{B}=1$ ($\lambda_\mathrm{B}=
1.12$ for the isothermal gas).

\begin{figure}
\vspace*{-0.5cm} 
\hspace*{-0.5cm} \centering
\includegraphics[width=9cm]{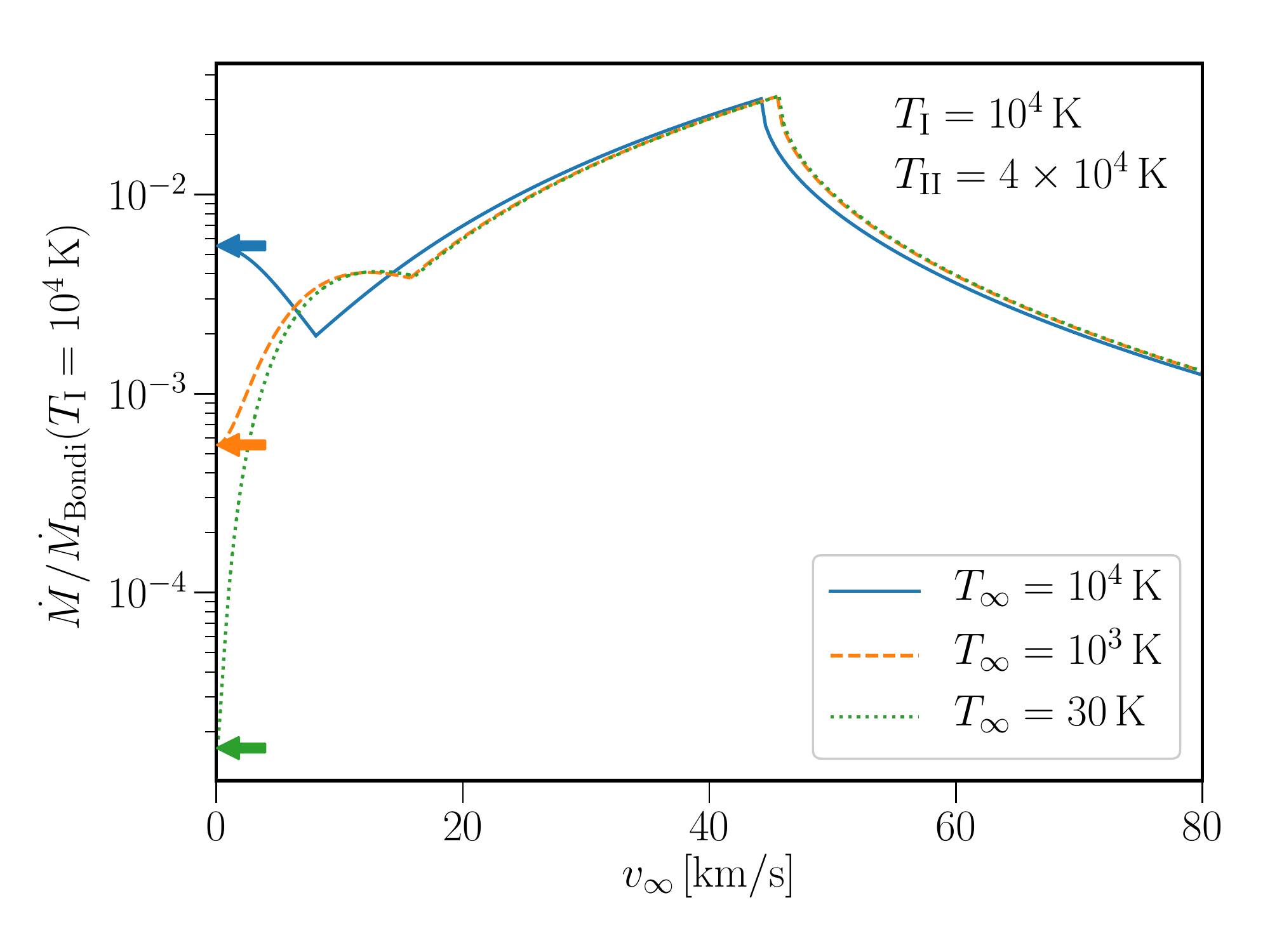} 
\vspace*{-0.5cm} 
\caption{Same as Fig.~\ref{fig:mdot_PR13} but for the accretion rate
normalized by the Bondi rate for neutral gas with
$T_\mathrm{I}=10^4\,\mathrm{K}$, $\dot{M}_\mathrm{Bondi}=4\pi G^2
M_\mathrm{BH}^2/c_\mathrm{I}^3$, for the velocity range including the
subsonic case.  We here plot the accretion rate for the case with the
constant temperature outside the I-front given by
Eqs.~\eqref{eq:mdot_model} and \eqref{eq:Mdot_sub} and those for the
case with X-ray preheating given by Eq.~\eqref{eq:mdot_model_xPH}. The
temperatures of ambient gas are $T_\infty=10^4\,\mathrm{K}$ (solid),
$10^3\,\mathrm{K}$ (dashed), and $30\,\mathrm{K}$ (dotted), while that
of the ionized region is $T_\mathrm{II}=4\times10^4\,\mathrm{K}$ and
that of the neutral region after X-ray preheating is
$T_\mathrm{I}=10^4\,\mathrm{K}$, if exists.  The arrows represent the
values for the static BHs.}  \label{fig:mdot_anl_extended}
\end{figure}

To make an interpolation between $v_\infty=0$ and $ c_\mathrm{I}$, we
assume that the velocity inside the ionized bubble scales linearly with
$v_\infty$ as $v_\mathrm{II}\approx
c_\mathrm{II}\,(v_\infty/c_\mathrm{I})$ and that the total pressure is
conserved along the flow, which gives the density
$\rho_\mathrm{II}=\rho_\infty(c_\mathrm{I}^2+v_\infty^2)/(c_\mathrm{II}^2+v_\mathrm{II}^2)$.
Then, the formula for the BHL accretion rate (the second expression in
Eq.~\ref{eq:mdot_model}) yields the accretion rate for $0\leq v_\infty <
c_\mathrm{I}$,
\begin{align}
 \dot{M}_\mathrm{sub-soinc} = 
\frac{4\pi G^2 M_\mathrm{BH}^2\rho_\infty\, \left(c_\mathrm{I}^2+v_\infty^2\right)}{c_\mathrm{II}^5}
\left(1+\left(\frac{v_\infty}{c_\mathrm{I}}\right)^2\right)^{-\frac{5}{2}}\,,
\label{eq:Mdot_sub}
\end{align}
which converges to $\dot{M}_\mathrm{v=0}$ at $v=0$ and
$\dot{M}_\mathrm{D}$ (Eq.~\ref{eq:mdotD}) at $v=c_\mathrm{I}$, as
expected.  In Fig.~\ref{fig:mdot_anl_extended}, we show $\dot{M}$ given
by Eqs.~\eqref{eq:mdot_model} and \eqref{eq:Mdot_sub} for $0\leq
v_\infty \leq 80\,\mathrm{km/s}$.

  \subsection{Case with X-ray preheating}
  \label{sec:X-ray-preheat}

Next, let us consider the case with X-ray preheating.  Although we
assume the constant neutral gas temperature $T_\mathrm{I}$ in
Sec.~\ref{sec:1d-model} and Appendix.~\ref{sec:subsonic-case}, the
temperature far from the BH can be lowered by the metal or molecular
cooling, as studied in \cite{Toyouchi:2020aa}. In their simulations with
metal cooling, the temperature of incident gas gradually increases as
the gas approaches the BH due to the heating by the X-rays leaking out
of the I-front (\ie, X-ray preheating).  Here, we assume that even if
the ambient temperature $T_\infty$ is cold, the incident flow is heated
by X-rays to a temperature $T_\mathrm{I}\sim 10^4\,\mathrm{K}$, which is
determined by the steep temperature dependence of the Ly$\alpha$
cooling, before reaching either the shock (in D-type flows) or the
I-front (in R-type flows).  In the following, we adopt subscripts
$\infty$ and $\mathrm{I}$ for quantities before and after the X-ray
preheating, respectively. The analytical model in
Sec.~\ref{sec:1d-model} and Appendix~\ref{sec:subsonic-case} is still
valid if we assume as boundary conditions at infinity the values after
the X-ray preheating.  Note that X-ray preheating may be suppressed due
to the attenuation of X-rays by gas or dust grains in the dense shell
preceding the D-type I-fronts. Using the effective cross-section for
Compton scattering and dust attenuation per hydrogen atom
$\sigma_\mathrm{eff}=\sigma_\mathrm{T}\,(1+710\,\,(Z/Z_\odot))$
\citep{Yajima:2017aa} and the column density of the shell $\Delta
N_\mathrm{shell}$ given by Eq.~\eqref{eq:DN_shell}, we can give a rough
estimate of the optical depth to X-rays of the shell:
\begin{align}
 \tau_\mathrm{X}&=\Delta
N_\mathrm{shell}\,\sigma_\mathrm{eff}\\ \nonumber
&=2\times
10^{-3}\,\left(\frac{n_\infty\,M_\mathrm{BH}}{10^6\,M_\odot\,\mathrm{cm^{-3}}}\right)
\,\left(1+710\,\,\left(\frac{Z}{Z_\odot}\right)\right)\, 
\left(1-\frac{v_\infty}{v_\mathrm{R}}\right)\,.
\end{align}
The equation above suggests that, assuming a relatively hard X-ray
spectrum, X-ray preheating can be suppressed in gas at solar or
supersolar metallicities, but is always effective in gas of primordial
composition. A soft X-rays spectrum could be more easily attenuated by
photoelectric absorption by helium and metal ions, perhaps suppressing
pre-heating even at relatively low gas metallicities. We plan to
investigate the effect of X-ray preheating in more detail in future
works.

In response to the temperature variation, the density and velocity also
change from their ambient values $\rho_\infty$ and $v_\infty$.  For the
supersonic part of the flow, we consider a quasi-1D plane-parallel model
similar to that for the R-type flows in Sec.~\ref{sec:1d-model}. From
mass and momentum conservation, we obtain the analogue of the I-front
jump condition given by Eq.~\eqref{eq:jump_IF}: the density
$\rho_\mathrm{I}$ and velocity $v_\mathrm{I}$ after the X-ray preheating
are given as $\rho_\mathrm{I}/\rho_\infty=v_\infty/v_\mathrm{I} =
\Delta^{(-)}(v_\infty,\,c_\infty,\,c_\mathrm{I})$ if
$v_\infty>v_\mathrm{R}(c_\infty,\,c_\mathrm{I})$, where
$v_\mathrm{R}(c_\infty,\,c_\mathrm{I}) \approx 2\,c_\mathrm{I}$ for
$c_\infty\ll c_\mathrm{I}$. If
$v_\infty<v_\mathrm{R}(c_\infty,\,c_\mathrm{I})$, however, a shock may
form in the pre-heating transition region because the above jump
conditions give unphysical (imaginary) values, as in the case of the
D-type flows in Sec.~\ref{sec:1d-model}.  In such a case, we again
interpolate the results for $v_\infty=0$ and
$v_\mathrm{R}(c_\infty,\,c_\mathrm{I})$, as the quasi-1D plane-parallel
approximation is not appropriate for the post-shocked subsonic flows.
Since $v_\mathrm{I}$ is equal to $0$ if $v_\infty=0$ and $c_\mathrm{I}$
if $v_\infty=v_\mathrm{R}(c_\infty,\,c_\mathrm{I})$, we assume
$v_\mathrm{I}=c_\mathrm{I}\,
v_\infty/v_\mathrm{R}(c_\infty,\,c_\mathrm{I})$.  We obtain the density
$\rho_\mathrm{I}=\rho_\infty(c_\infty^2+v_\infty^2)/(c_\mathrm{I}^2+v_\mathrm{I}^2)$
from the conservation of the total pressure, as assumed in
Sec.~\ref{sec:1d-model}.

Finally, we obtain the accretion rate $\dot{M}$ for the case with X-ray
preheating, using the above expressions for $\rho_\mathrm{I}$ and
$v_\mathrm{I}$ after the temperature reaches $T_\mathrm{I}$. From the
formula for $\dot{M}$ for the case with fixed neutral-gas temperature
(Eqs.~\ref{eq:mdot_model} and \ref{eq:Mdot_sub}), we obtain
\begin{align}
 \dot{M} = 
\begin{cases}
\dot{M}_\mathrm{sub-sonic} &  0 \leq v_\infty < v_\mathrm{R}(c_\infty,\,c_\mathrm{I})\\
\dot{M}_\mathrm{D} &  v_\mathrm{R}(c_\infty,\,c_\mathrm{I}) < v_\infty < v_\mathrm{R}(c_\mathrm{I},\,c_\mathrm{II})\\
\dot{M}_\mathrm{R} & v_\mathrm{R}(c_\mathrm{I},\,c_\mathrm{II}) < v_\infty 
\end{cases}
\,,
 \label{eq:mdot_model_xPH}
\end{align}
with
\begin{align}
 \dot{M}_\mathrm{sub-soinc} = 
\frac{4\pi G^2 M_\mathrm{BH}^2\rho_\infty \left(c_\infty^2+v_\infty^2\right)}{c_\mathrm{II}^5}
\left(1+\left(\frac{v_\infty}{v_\mathrm{R}(c_\infty,\,c_\mathrm{I})}\right)^2\right)^{-\frac{5}{2}}\,,
\label{eq:Mdot_sub_xPH}
\end{align}
\begin{align}
\dot{M}_\mathrm{D}= \frac{\pi G^2 M_\mathrm{BH}^2\rho_\infty\, (v_\infty^2 + c_\infty^2)}{\sqrt{2}c_\mathrm{II}^5}\,,
\label{eq:mdotD_xPH}
\end{align}
and
\begin{align}
\dot{M}_\mathrm{R}=\frac{4\pi G^2 M_\mathrm{BH}^2\rho_\infty\Delta^{(-)}(v_\infty,c_\infty,c_\mathrm{II})}
 {\left(c_\mathrm{II}^2+\left(\frac{v_\infty}{\Delta^{(-)}(v_\infty,c_\infty,c_\mathrm{II})}\right)^{2}\right)^\frac{3}{2}}\,.
\label{eq:mdotR_xPH}
\end{align}
In Fig.~\ref{fig:mdot_anl_extended}, we plot $\dot{M}$ for the case with
X-ray preheating, setting $T_\infty=30$ and $10^3\,\mathrm{K}$ and
$T_\mathrm{I}=10^4\,\mathrm{K}$.  In Eqs.~\eqref{eq:mdotD_xPH} and
\eqref{eq:mdotR_xPH}, we have simplified the expressions using the
conservation of the total pressure and the mass flux along the flow.
Note that the velocities at which the expression for $\dot{M}$ changes
from $\dot{M}_\mathrm{sub-sonic}$ to $\dot{M}_\mathrm{D}$ and from
$\dot{M}_\mathrm{D}$ to $\dot{M}_\mathrm{R}$ are
$v_\mathrm{R}(c_\infty,\,c_\mathrm{I})\approx 2\,c_\mathrm{I}$
$(c_\infty\ll c_\mathrm{I})$ and
$v_\mathrm{R}(c_\mathrm{I},\,c_\mathrm{II})\approx 2\,c_\mathrm{II}$
$(c_\mathrm{I}\ll c_\mathrm{II}$), respectively.  The accretion rate
converges to the ordinary BHL value for $v_\infty \gg 2\,c_\mathrm{II}$.

     \bsp	
     \label{lastpage}
  \end{document}